\documentclass[twocolumn]{aastex6}

\usepackage{natbib}
\usepackage{amsmath}
\usepackage{multirow}
\usepackage{graphicx}
\usepackage{color}
\usepackage{threeparttable}

\def\gx339{GX~339--4}

\def\rxte{{\it RXTE}}
\def\xmm{{\it XMM-Newton}}
\def\suzaku{{\it Suzaku}}
\def\swift{{\it Swift}}
\def\nustar{{\it NuSTAR}}

\def\xillver{{\tt xillver}}

\def\xillverCp{{\tt xillverCp}}
\def\relxill{{\tt relxill}}

\def\relxillCp{{\tt relxillCp}}

\def\relline{{\tt relline}}
\def\nthComp{{\tt nthComp}}
\def\simplcut{{\tt simplcut}}
\def\diskbb{{\tt diskbb}}

\begin{document}
\title{The evolution of \gx339\ in the low-hard state as seen by
\textit{N\MakeLowercase{u}STAR} and \textit{S\MakeLowercase{wift}}}
\author{Jingyi~Wang-Ji\altaffilmark{1},
Javier~A.~Garc{\'\i}a \altaffilmark{2,3},
James~F.~Steiner\altaffilmark{4},
John~A.~Tomsick\altaffilmark{5},
Fiona~A.~Harrison\altaffilmark{2},
Cosimo~Bambi\altaffilmark{1,6},
Pierre-Olivier~Petrucci\altaffilmark{7},
Jonathan~Ferreira\altaffilmark{7},
Susmita~Chakravorty\altaffilmark{8} and 
Ma\"{i}ca~Clavel\altaffilmark{5}.}
\affil{
\altaffilmark{1}Center for Field Theory and Particle Physics, Department of
Physics, Fudan University, 200433 Shanghai, China\\ 
\altaffilmark{2}Cahill Center for Astronomy and Astrophysics, California
Institute of Technology, Pasadena, CA 91125, USA; javier@caltech.edu\\ 
\altaffilmark{3}Remeis Observatory \& ECAP, Universit\"{a}t
Erlangen-N\"{u}rnberg, 96049 Bamberg, Germany\\ 
\altaffilmark{4}MIT Kavli Institute for Astrophysics and Space
Research, MIT, 70 Vassar Street, Cambridge, MA 02139, USA\\
\altaffilmark{5}Space Sciences Laboratory, University of California, Berkeley,
CA 94720, USA\\ 
\altaffilmark{6}Theoretical Astrophysics, Eberhard-Karls Universit\"at T\"ubingen, 72076 T\"ubingen, Germany\\
\altaffilmark{7}Universit\'{e} Grenoble Alpes, CNRS, IPAG, F-38000 Grenoble,
France\\ 
\altaffilmark{8}Department of Physics, Indian Institute of Science, Bangalore
560012, India.}

\begin{abstract}
We analyze eleven \nustar\ and \swift\ observations of the black
hole X-ray binary \gx339\ in the hard state, six of which were taken during
the end of the 2015 outburst, five during a failed outburst in 2013. These
observations cover luminosities from $0.5\%-5\%$ of the Eddington luminosity.
Implementing the most recent version of the reflection model
\texttt{relxillCp}, we perform simultaneous spectral fits on both datasets to
track the evolution of the properties in the accretion disk including the inner
edge radius, the ionization, and temperature of the thermal emission.
We also constrain the photon index and electron temperature of the primary
source (the ``corona"). We observe a maximum truncation radius of $37$~$R_g$ in the preferred fit for the 2013 dataset, and a marginal correlation between the level of truncation and luminosity. We also explore a self-consistent model under the framework of coronal
Comptonization, and find consistent results regarding the disk truncation in
the 2015 data, providing a more physical preferred fit for the 2013
observations. 
\end{abstract}

\section{Introduction} \label{introduction}

\gx339\ is a low mass X-ray binary (LMXB) and an archetypical black hole
transient that shows a high level of activity in optical, infrared, radio and
X-rays, with more than a dozen outburst cycles (typically every 2--3 years) of
different strengths since its first discovery in 1973 \citep{markert73}. The
high flux it can achieve in the hard state and the recurrent outburst activity
make \gx339\ an ideal source to study the evolution of the accretion disk in
the low-hard state. A recent near-infrared study in \citet{heida2017potential}
has shown a mass function of $1.91\pm0.08$~$M_{\odot}$, much less than previously
claimed ($5.8\pm0.5$~$M_{\odot}$, \citealt{hynes2003dynamical}); the inclination
angle of the system is $37\degr<i<78\degr$ from optical analysis, and the black
hole mass can be as small as 2.3~$M_\odot$ with 95\% confidence.  
%-+

The evolution of the accretion disk properties is an observational foundation
essential to understand the physics governing the outbursts of LMXB systems.
A body of evidence has shown that when a black hole binary is in the soft
state, the accretion disk extends to the innermost stable circular orbit (ISCO,
e.g.; \citealt{steiner2010constant,gierlinski2004black}). The standard
paradigm for the low-hard state is that the disk's truncation radius grows as
luminosity decreases, leaving an interior hot advection-dominated accretion
flow (ADAF, \citealt{narayan2008advection}) or other coronal flow (e.g.;
\citealt{ferreira2006unified}). There is good evidence that at very low
luminosities the disk is largely truncated (see \citealt{narayan2008advection}
for a review). However, for luminosities in a moderate range of $0.1-10\%$ of
the Eddington limit, the values of reported inner edge of the disk ($R_{in}$)
vary significantly, making this a hotly debated topic. There are two widely
adopted methods to estimate $R_{in}$: the continuum-fitting method, focusing on
the thermal emission of the disk; and the reflection spectroscopy (commonly
called the iron-line method), which models the reflection component coming from
the reprocessing of the Comptonized photons illuminating the optically-thick
disk. In this paper we make use of the latter, since our observations are in
the low-hard state, where the hard continuum and the reflected components
dominate the spectra.
%-+

The reflection spectrum is a rich mixture of radiative recombination continua,
absorption edges, fluorescent lines (most notably the Fe K complex in the $6-8$
keV energy range), and a Compton hump at energies $>10$ keV. This reflected radiation
leaves the disk carrying information on the physical composition and condition
of the matter in the strong gravitational field near the black hole. The
fluorescent lines are broadened and shaped by Doppler effects, light bending
and gravitational redshift. Under the assumption that astrophysical black
holes are Kerr black holes, the method can be used to measure the spin
parameter $a_*=cJ/GM^2$ ($-1\leq a_* \leq 1$), where $J$ is the black hole spin angular momentum and $M$ is the black hole mass. By estimating the radius of the
inner edge of the accretion disk, so long as the inner radius corresponds to
the radius of the innermost stable circular orbit, $R_{ISCO}$, which simply and
monotonically maps to $a_*$ \citep{hughes2003black}, we can measure the black hole spin. For the three canonical
values of the spin parameter, $a_*=+1$, 0 and --1, $R_{ISCO}=1M$, $6M$ and $9M$
($c=G=1$). Alternatively, by fixing the spin parameter to its maximal value in \relxill\
($a_*=0.998$), one can estimate the maximally truncation of the inner radius of
the disk. 
%-+

The most advanced reflection model to date is
\relxill\ \citep{garcia2014improved,dauser2014role}, which is based on
the reflection code \xillver\ \citep{garcia2010,garcia2013x}, and the
relativistic line-emission code \relline\
\citep{dauser2010broad,dauser2013irradiation}. The \relxill\ model
family has different flavors\footnote{\url{www.sternwarte.uni-erlangen.de/research/relxill}}.  In two of
these, the modeling of the incident spectrum is done by either the standard
power law with a high-energy cutoff in the form of an exponential rollover, or
by the continuum produced by a thermal Comptonization model (\nthComp,
\citet{zdziarski1996broad}). The results presented in this paper are derived
using \relxillCp\ to model the relativistically-blurred reflection
component from the inner disk and \xillverCp\ to model unblurred
reflection from a distant reflector, both adopting the continuum produced by the \nthComp\ model. 
%-+

In the past ten years, great effort has been devoted to estimate the inner edge
of the accretion disk of \gx339\ in the low-hard state with reflection
spectroscopy, analysing data from 8 outburst cycles of \gx339\ (2002, 2004,
2007, 2008, 2009, 2010-2011, 2013, 2015) obtained from X-ray missions including
\xmm\
\citep{miller2006long,reis2008systematic,kolehmainen2013soft,plant2015truncated,basak2016spectral},
\swift\ \citep{tomsick2008broadband}, \suzaku\
\citep{tomsick2009truncation,shidatsu2011x,petrucci2014return}, {\it Rossi
X-ray Timing Explorer} ({\it RXTE}, \citealt{javier_gx339}), the
{\it Nuclear Spectroscopic Telescope Array} ({\it NuSTAR},
\citealt{furst2015complex}). 
%-+

Analyzing \xmm\ data with reflection spectroscopy, \citet{miller2006long}
presented for the first time strong evidence that the disk extended closely to
the ISCO ($R_{in}=5\pm0.5 R_g$) in the bright phase of the low-hard state
($L/L_{edd}\sim 5.4\%$ assuming $M_{bh}=10$~$M_\odot$ and $D=8$ kpc), which was
later confirmed by \citet{reis2008systematic} using the same \xmm\ EPIC-MOS
data taken in 2004. These results were challenged by \citet{done2010re}, who
reported that the iron line profile appears much narrower in the \xmm\ EPIC-pn
data taken in timing mode (for the same observation), presumably because in
this mode the pile-up is reduced.  They obtained a much larger disk truncation
($R_{in}=60^{+40}_{-20} R_g$).  Other authors have also reported large disk
truncation by analyzing the same EPIC-pn timing mode data:
$R_{in}=115^{+85}_{-35} R_g$ \citep{kolehmainen2013soft},
$R_{in}=316^{+164}_{-74} R_g$ \citep{plant2015truncated},
$R_{in}=227^{+211}_{-84} R_g$ and $144^{+107}_{-96} R_g$ separating the two
revolutions \citep{basak2016spectral}. Nevertheless,
\citealt{miller2010relativistic} argued that pile-up can still affect the
timing mode, and that if not corrected it can artificially make the continuum
softer, which in turn will result in a narrower Fe K profile, leading to false
estimates of large truncation. The discussion centered around pile-up effects
suggest that it is a complicated instrumental issue for X-ray charge-coupled
devices (CCD), for which we still do not have a complete model.
%-+

\citet{javier_gx339} have independently analyzed the \rxte/PCA data tracking
the evolution of \gx339\ in the hard state with the luminosity ranging from
17\% to 2\% of the Eddington luminosity. Although the PCA data do not have
problems with photon pile-up, and has archived extremely high signal-to-noise
ratio and low systematic uncertainty by implementing the \texttt{PCACORR} tool
\citep{garcia2014_pcacorr}, it is limited by its relatively low spectral
resolution to study the iron line complex. With the most recently available
data from \nustar\ (which is also free from pile-up), we can now extend the
luminosity range down to $0.5\% L_{edd}$, to see the evolution of the
accretion disk's truncation and other conditions in the system.
%-+

In this paper, we focus on \swift\ and \nustar\ to sidestep pile-up issues
noting that there is some disagreement between the \xmm\ and \nustar\ spectra.
For example, in the recent analysis presented by \citealt{stiele2017nustar},
the \nustar\ spectra can only be used down to 4~keV (see Figure~7 therein) due
to this discrepancy. Thus, since the combination of \xmm\ and \nustar\
observations seems to require a special treatment, a detailed analysis of such
data will be presented in a future publication. 
%-+

This paper is organized as follows. Section~\ref{obs} describes the
observations and data reduction, Section~\ref{fitting} provides the details of
our spectral fitting. We present our discussion in Section~\ref{dis}, and
summarize the results in Section~\ref{conclusion}.
%-+

\section{Observations and data reduction}\label{obs}

\begin{figure}[htb!]
\centering
\includegraphics[width=\linewidth]{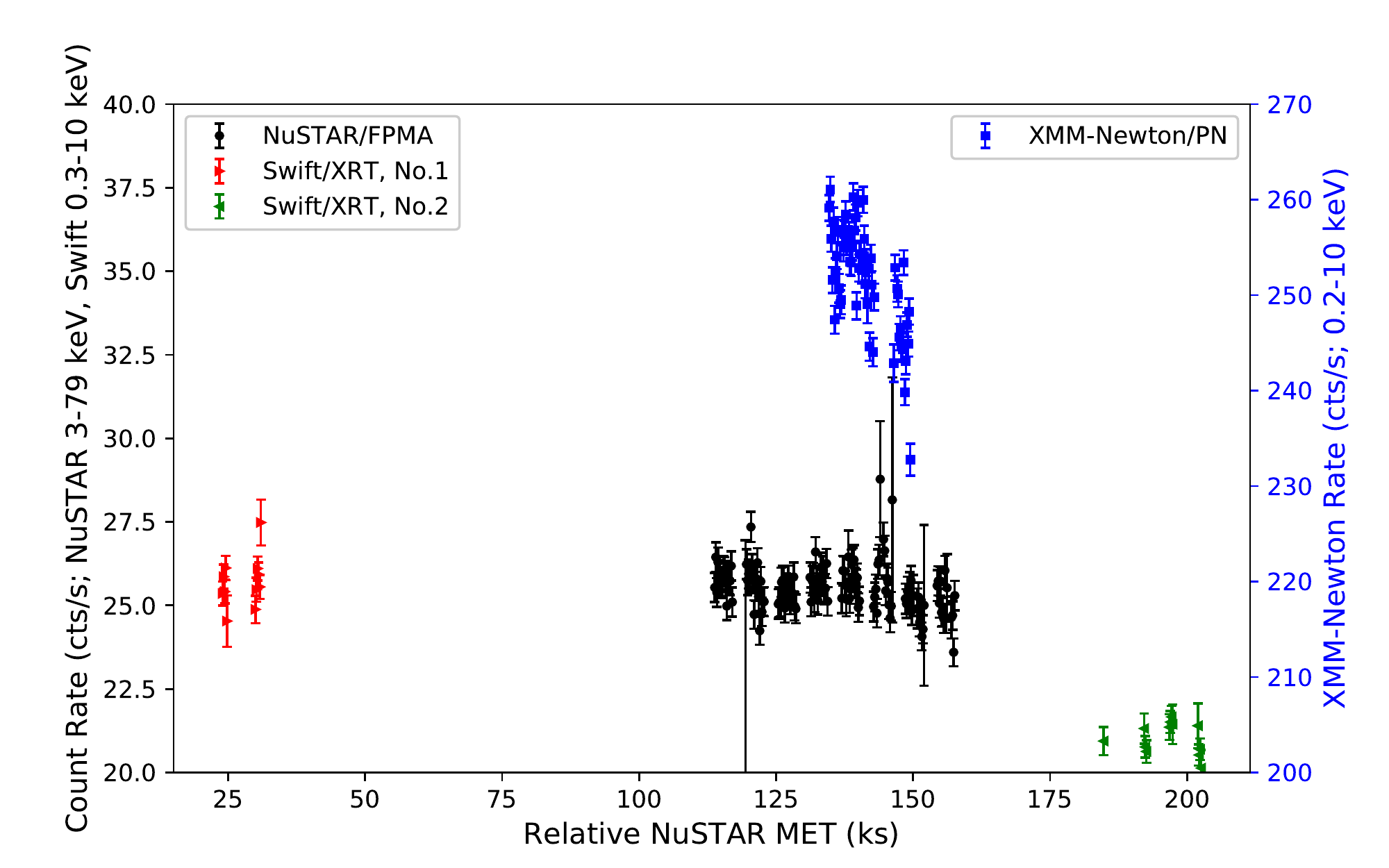}
\caption{
Light curves of \swift/XRT (0.3-10 keV), \nustar/FPMA (3.0-79.0 keV), and
\xmm/EPIC-PN (0.2-10 keV) during the first observation in 2015, bin-time: 200
s. \swift/XRT, No.1 and No.2  refer to \swift\ obs.ID of 00032898123 and
00032898124, which were both taken within a day from \nustar's start time. The
reference time is $1.7835\times10^8$s (\nustar\ MET, also 2015-08-27 05:39:58
UTC or 57261 MJD).
}
\label{fig:lc}
\end{figure}

In August 2015, X-ray monitoring detected the end of a new outburst of \gx339\
and triggered observations with the \nustar\ \citep{harrison2013nuclear},
\swift\ \citep{gehrels2004swift} and \xmm\ \citep{jansen2001xmm}. We obtained
six observations with \nustar\ at the end of the outburst, and for each a
corresponding \swift\ snapshot within a day of the start time of \nustar\
(Figure~\ref{fig:lc}). 
%-+

We also analyzed the dataset from 2013, which was triggered by the detection of
the onset of a new outburst. In this campaign five observations were taken with
\nustar, four during the rise and one during the decay of the outburst, and
\swift\ observations every other day. However, the 2013 was a failed outburst because 
the source did not follow the standard outburst pattern in the hardness-intensity
diagram. The source remained in the low-hard state, and never switched to the
high-soft state \citep{furst2015complex}. Table~\ref{tab:obs} provides a
detailed observation log of the \nustar\ and the matching \swift\ observations.
%-+

%-+-------------------------------------------------------------------------------
%-+-------------------------------------------------------------------------------

%-+-------------------------------------------------------------------------------
\begin{table*}
\begin{center}
\caption{\nustar\ and \swift\ observations in the 2015 and 2013 outburst
cycles, exposure times and start time. WT: windowed timing mode, PC: photon
counting mode. \label{tab:obs}}
\footnotesize
\begin{tabular}{cccc|ccc|cccc}\hline \hline
outburst&No.&$F_{2-10 keV}$&$L/L_{edd}$&\multicolumn{3}{c|}{\texttt{NuSTAR}}&\multicolumn{4}{c}{\texttt{Swift}}\\
&&($10^{-10}$ergs/cm$^2$s)&(\%)&obs.ID&S.T.&exp.(ks)&obs.ID&S.T.&exp.(ks)&mode\\
\hline
2015&1&6.92&2.0&80102011002&08-28 13:06&21.6&00032898124&08-29 08:55&1.7&WT\\
&2&5.64&1.8&80102011004&09-02 12:36&18.3&00032898126&09-03 00:37&2.3&WT\\
&3&4.77&1.7&80102011006&09-07 14:51&19.8&00032898130&09-07 00:21&2.8&WT\\
&4&3.66&1.2&80102011008&09-12 15:46&21.5&00081534001&09-12 16:18&2.0&WT\\
&5&2.54&1.0&80102011010&09-17 10:06&38.5&00032898138&09-17 00:06&2.3&WT\\
&6&1.32&0.5&80102011012&09-30 01:11&41.3&00081534005&09-30 05:32&2.0&PC\\
\hline
2013&1&3.44&1.4&80001013002&08-11 23:46&42.3&00032490015&08-12 00:33&1.1&WT\\
&2&5.68&2.4&80001013004&08-16 17:01&47.4&00080180001&08-16 18:22&1.9&WT\\
&3&8.70&3.6&80001013006&08-24 12:36&43.4&00080180002&08-24 04:02&1.6&WT\\
&4&11.85&4.6&80001013008&09-03 09:56&61.9&00032898013&09-02 19:03&2.0&WT\\
&5&2.06&0.8&80001013010&10-16 23:51&98.2&00032988001&10-17 11:57&9.6&WT\\
\hline
\end{tabular}
\raggedright{\textbf{Notes.} \\Luminosity calculated using unabsorbed flux
between $0.1-300$ keV, assuming a distance of 8 kpc and a black hole mass of 10~
$M_{\odot}$.}
\end{center}
\end{table*}
%-+-------------------------------------------------------------------------------

\subsection{NuSTAR}

The \nustar\ data were reduced using the Data Analysis Software (NUSTARDAS)
1.7.1, which is part of HEASOFT~6.21 and CALDB version 20170614.  Source
spectra were extracted from 100$''$ circular extraction regions centered on the
source position, and background spectra from 135$''$ circular regions from the
opposite corner of the detector. We binned the spectra from \nustar's focal point
modules A and B (FPMA and FPMB) to oversample the spectral resolution by a
factor of 3, to 1 minimal count per bin for C-statistics. We fitted the spectra
over the whole energy range (3--79~keV) using the C-statistics.
%-+

\subsection{Swift}

The \textit{Swift}/XRT data were processed with standard procedures
(\texttt{xrtpipeline 0.13.3}), filtering and screening criteria using FTOOLS
6.21. The data collected in the windowed timing mode were not affected by
pile-up, so source events were accumulated within a circle with the radius of 20
pixels (1 pixel $\sim$ 2.36$''$), background events within an annular region with
an outer radius of 110 pixels and inner radius of 90 pixels. For the last 2015
data collected in the photon counting mode, pile-up problem is a concern, so we
fitted the PSF profile with a King function in the wings, then extrapolated to
the inner region and saw the divergence resulting from pile-up. 
We accordingly excluded a circular region with radius of 5 pixels from the
source extraction region. For the response matrix, we used the response files
swxwt0to2s6\_20131212v015.rmf, swxwt0to2s6\_20130101v015.rmf for the
observations in 2015 and 2013, respectively. We generated the ancillary response
files including a correction using the exposure maps, accounting for the
effective area by \texttt{xrtmkarf}. The XRT spectra were rebinned also to 1
minimal count per bin. The fitted energy range is 0.5--8~keV.
%-+

All the uncertaintites quoted in this paper are for a 90\% confidence range,
unless otherwise stated. All spectral fitting is done with {\sc xspec}~12.9.1
\citep{arn96}.  In all the fits we use {\it wilm} set of abundances
\citep{wilms2000absorption}, and {\it vern} photoelectric cross sections
\citep{verner1996atomic}.
%-+

\section{Spectral fitting}\label{fitting}

%-+-------------------------------------------------------------------------------
\begin{figure*}[htb!]
\centering
\epsscale{1.1}
\plottwo{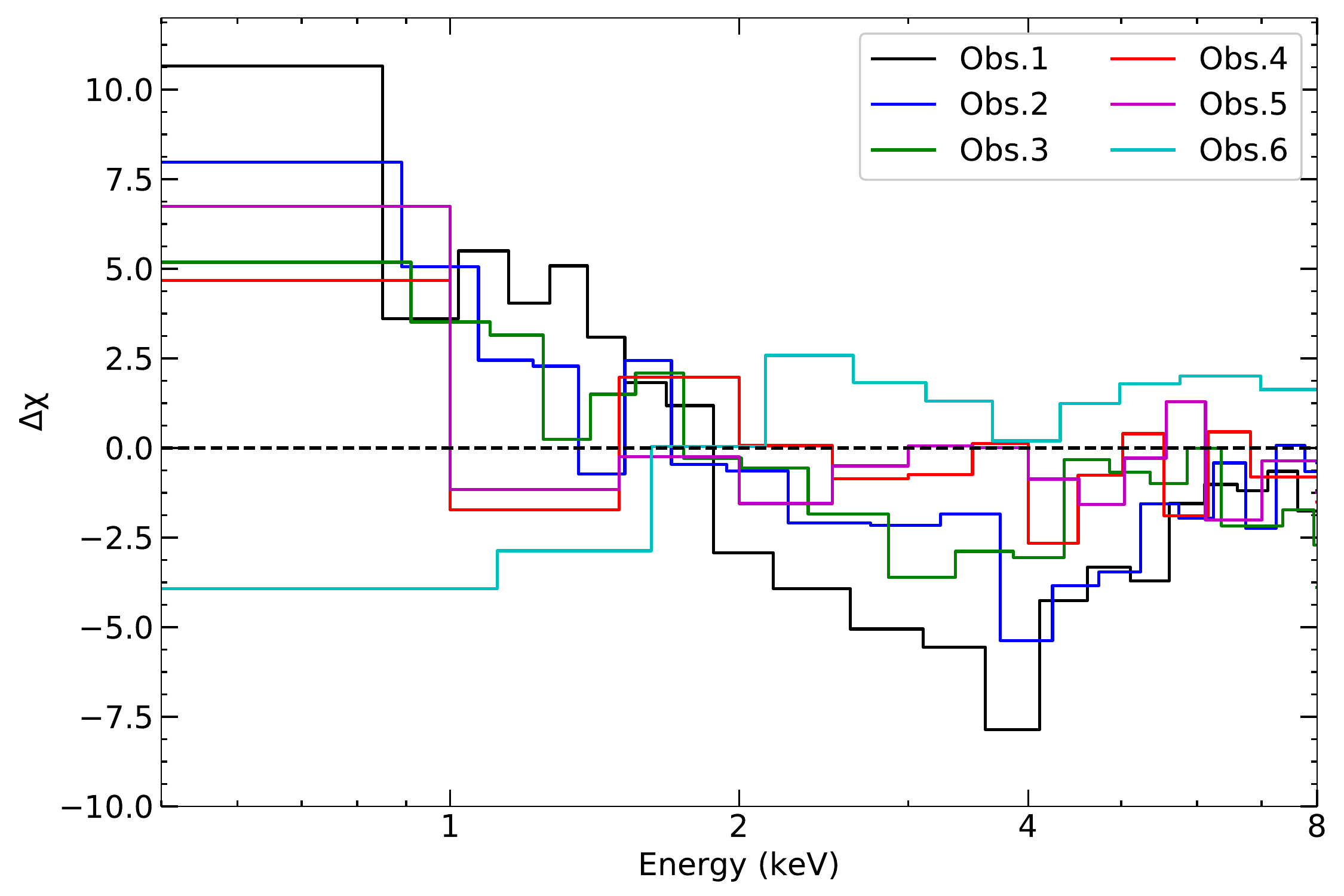}{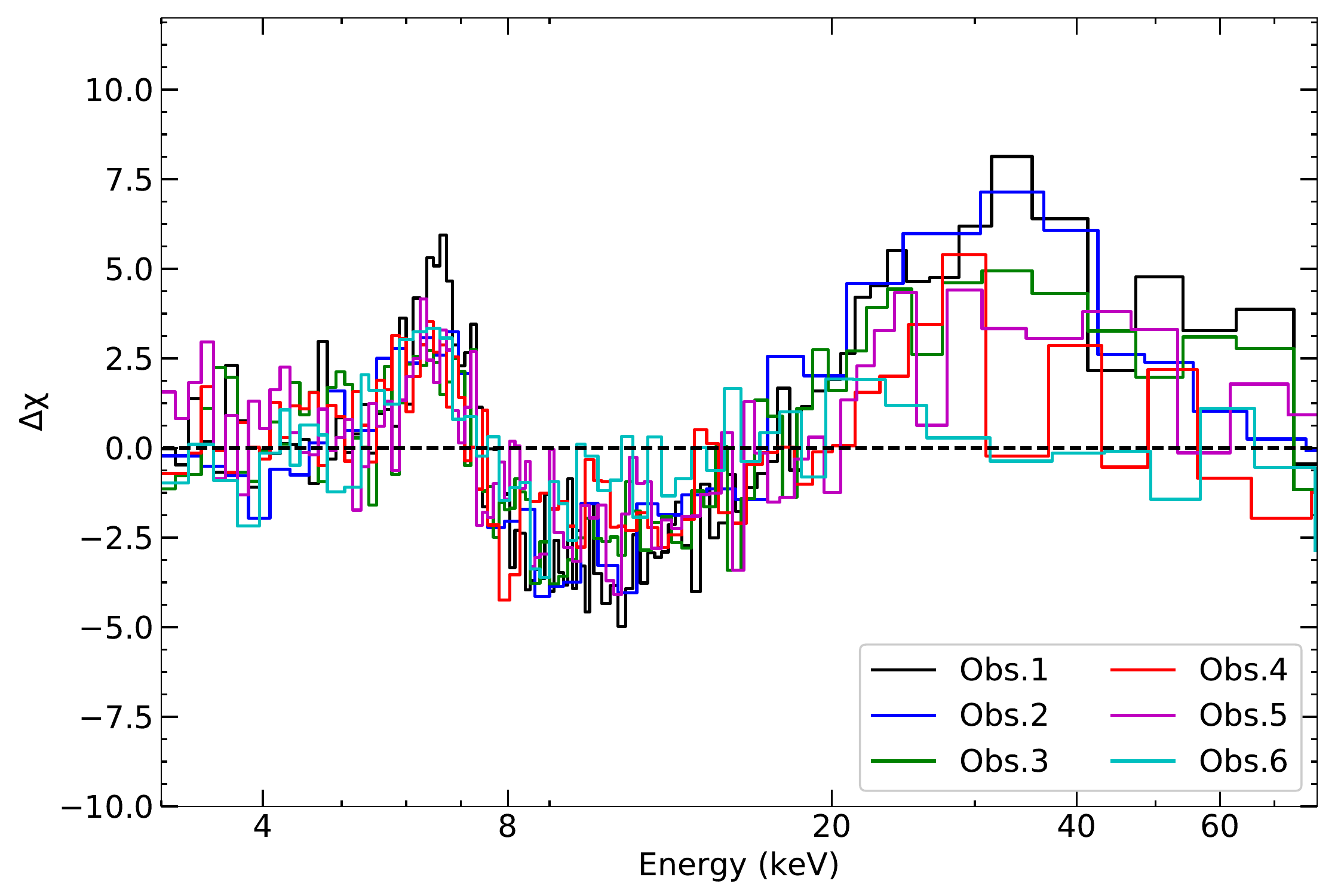}
\caption{$\Delta\chi$ for a fit with an absorbed power-law model (i.e.,
\texttt{tbabs*powerlaw}) in the $3-8$~keV range with a fixed column density $N_H=5\times
10^{21}$ cm$^{-2}$ for the 2015 dataset. The disk component is present with
good statistical precision in the first three observations from the \swift/XRT
part (\textit{left}), the iron line and Compton hump are clearly visible in all
observations from the \nustar\ part (\textit{right}, only FPMA data are plotted
here). Data are rebinned for display clarity. \label{fig:powerlaw}}
\end{figure*}
%-+-------------------------------------------------------------------------------

\subsection{The 2015 dataset: during decay in the hard state}

\subsubsection{Model 1: the standard reflection model}

After fitting with an absorbed power-law (i.e., \texttt{tbabs*powerlaw}) in the
$3-8$~keV range with a fixed column density $N_H=5\times 10^{21}$ cm$^{-2}$, we can
see from the the data-to-model ratio (Figure \ref{fig:powerlaw}) a disk
component at $\lesssim1-2$~keV is present except for the last observation, and
the iron line and Compton hump are clearly visible in all observations. Note
that the total number of counts in \swift\ drops dramatically from $\sim32,000$ counts
(observation 3) to $\sim4,900$ counts (observation 4), $\sim5,300$ counts
(observation 5) and $\sim1,500$ counts (observation 6), so the statistical
precision for the last three observations is relatively poor.
%-+

We perform a simultaneous fit on all six observations from 2015 using a more
sophisticated model: {\tt const*Tbabs*(diskbb+nthComp+relxillCp+xillverCp)}
(2015-M1), where \relxillCp\ models the relativistic reflection
component and \xillverCp\ represents the unblurred reflection coming
from a distant reflection that could be wind or the outer region of a flared
disk. The multi-color blackbody emission from the accretion disk is included
via \diskbb, and the Comptonization of the disk emission coming from
the corona via \nthComp. During the fit, we tie several global
parameters that are expected to be unchanged during the time range for our
observations ($\sim$ a month) including the column density $N_H$, the
inclination angle $i$, and the iron abundance $A_{Fe}$. The spin parameter $a_*$ is
fixed at its maximal allowed value of 0.998, while the inner radius is left free
to vary, so that $R_{in}$ can be fully explored. The constants are
introduced as cross-calibration factors, thus are frozen at 1.0 for FPMA, tied
together for all FPMB spectra but allowed to vary for XRT to account for the
possible differences in the flux levels since these observations are not
strictly simultaneous. The reflection fractions for the blurred and unblurred 
reflection components are frozen at $R_f=-1$, their iron abundances are tied,
and the ionization parameter is fixed at $log\xi=0$ in \xillverCp\, as the
gas in the distant reflector is expected to be cold and neutral (following
\citealt{javier_gx339}). The seed photon temperature $kT_{bb}$ in \texttt{nthcomp} is tied with the temperature at inner disk radius $kT_{in}$ in \texttt{diskbb}. If not specified, we use a canonical emissivity profile of
$\propto r^{-3}$ (i.e., emissivity index $q=3$). 
%-+

\begin{figure*}[htb!]
\plottwo{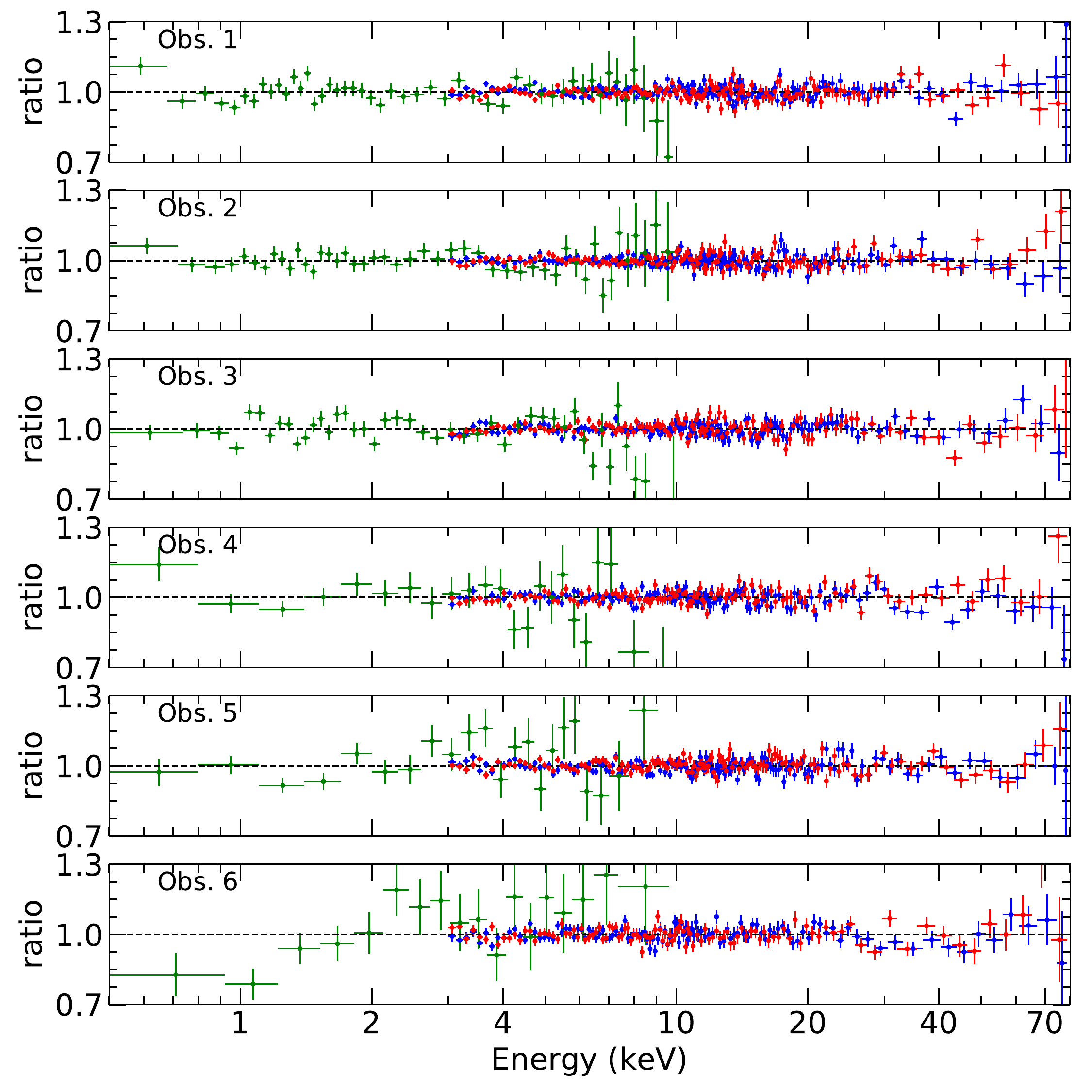}{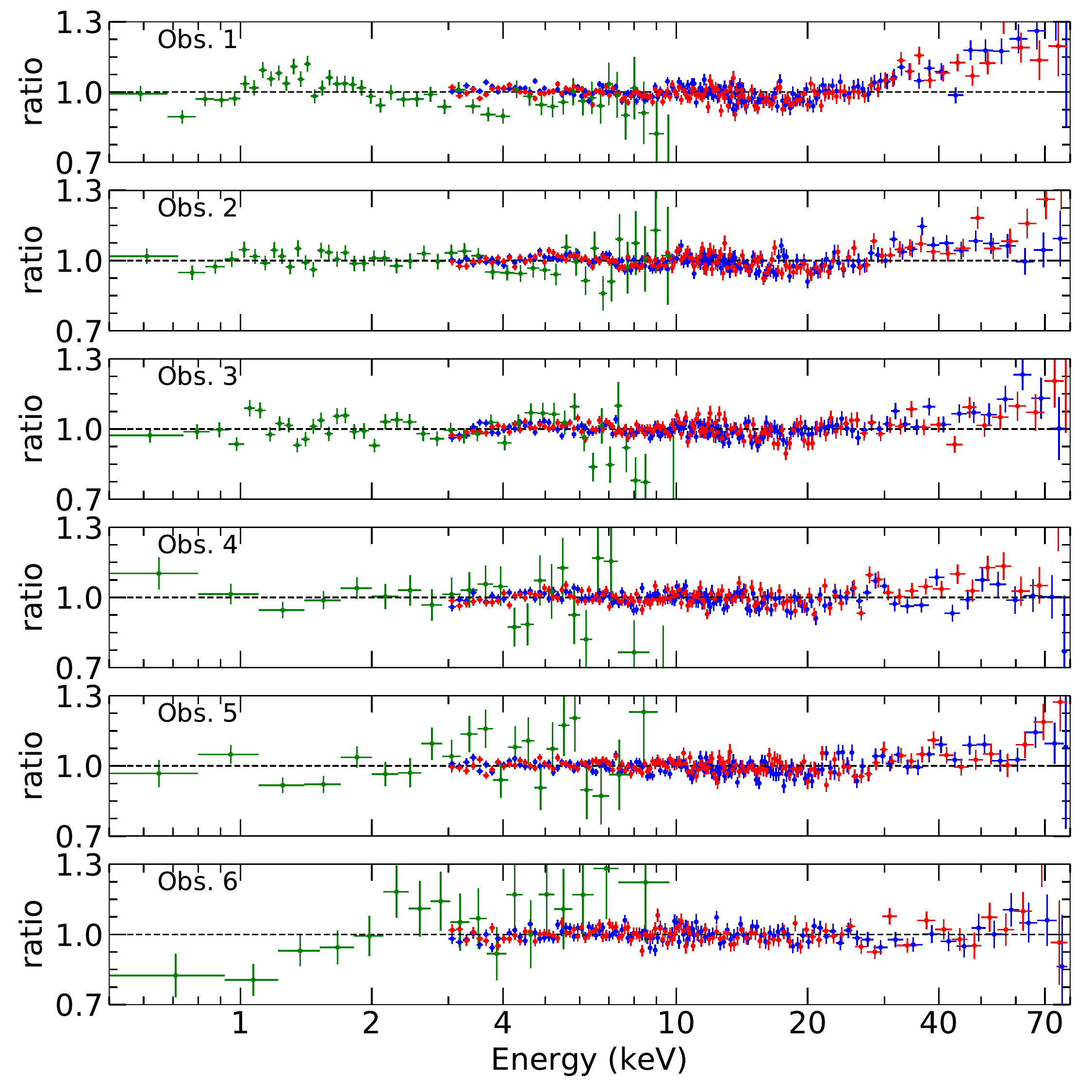}
\caption{Data-to-model ratio for the simultaneous fit with M1 performed on the
2015 dataset with free iron abundance (\textit{left}) and when the iron
abundance is fixed to be the solar value (\textit{right}). Discrepancy can be seen above $\sim
30$~keV when the iron abundance is fixed at the Solar value. 
This demonstrates the preference of these data to require large iron abundance.\label{fig:M1_cal_r}}
\end{figure*}

\begin{figure*}
\epsscale{1.15}
\plottwo{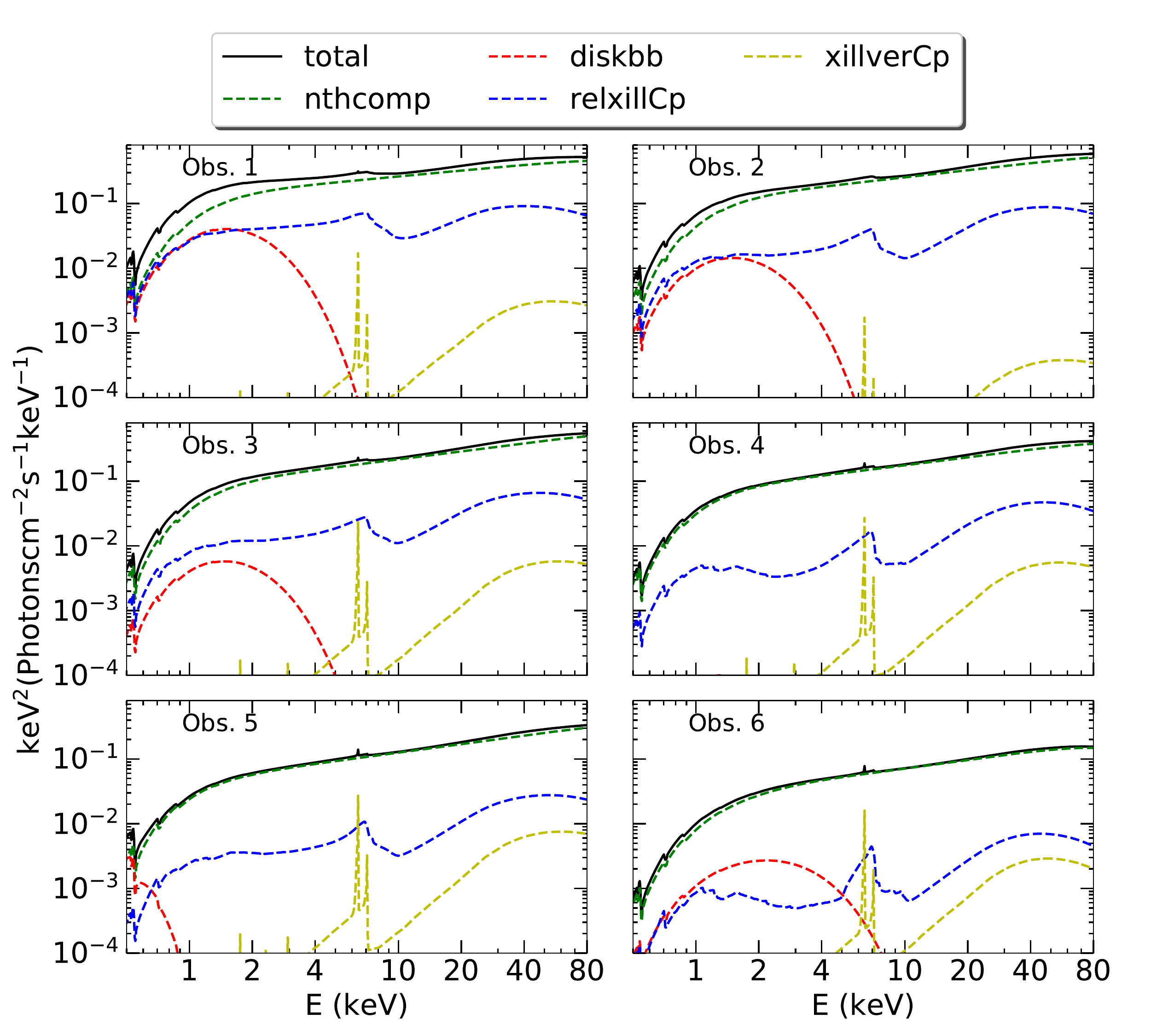}{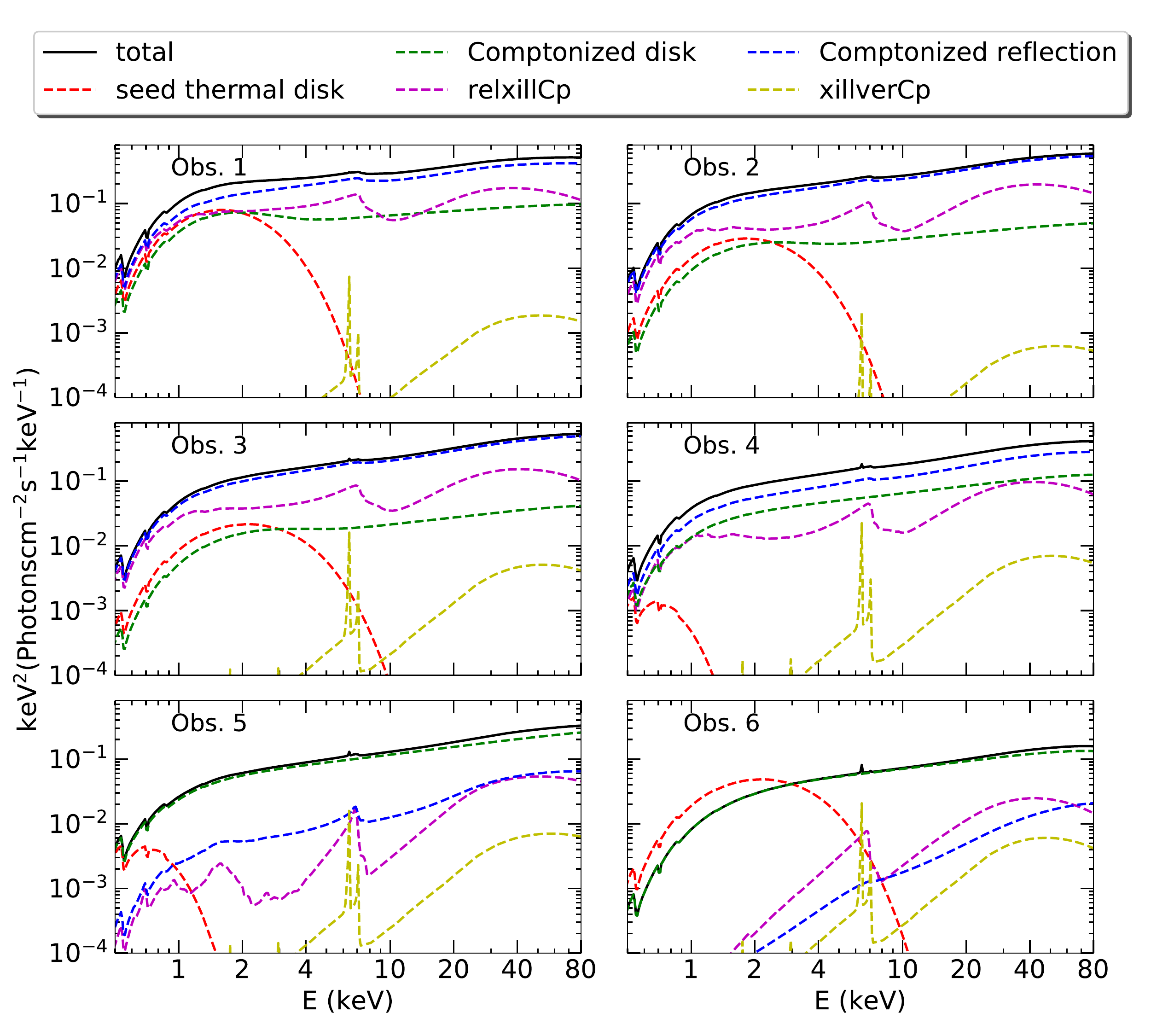}
%\caption{Model components for individual observations in 2015 for M1 ({\it left}) and M2 ({\it right}). ({\it left}) Black: total, red: \diskbb, green: \nthComp, blue: \relxillCp, yellow: \xillverCp. ({\it right}) Black: total, red: the intrinsic disk emission (\diskbb), magenta: the reflection component (\relxillCp), green: Comptonized disk including the unscattered intrinsic disk emission and the continuum component it generates (\simplcut\ convolved with \diskbb), blue: Comptonized reflection including the unscattered reflection and the Comptonized component it generates (\simplcut\ convolved with \relxillCp), yellow: \xillverCp.}
\caption{Model components for individual observations in 2015 for M1 ({\it left}) and M2 ({\it right}). The component each color represents is indicated in the figure. The spectrum becomes harder with the photon index dropping from 1.72 to 1.62 when the luminosity decreases. Although the statistical precision for the last three observations is relatively poor, a tentative decreasing trend in the disk temperature and the flux ratio between 2 and 20 keV of the disk component and the unabsorbed total one are shown except for the last observation.}
\label{fig:2015_eem}
\end{figure*}

The resulting ratio is shown in Figure \ref{fig:M1_cal_r} (\textit{left}), the
best fit parameter values in Table \ref{tab:2015_M1_cal} and the model
components in Figure \ref{fig:2015_eem} (\textit{left}). As we expect from the
dramatic drop in count number for the last three observations, the \swift\ data
can not provide solid constraints on the intrinsic disk emission. However, we
do obtain a decreasing trend in the disk temperature and the flux ratio between
2 and 20 keV of the disk component and the unabsorbed total one, except for the
last observation which has a physically unreasonable high disk temperature
$0.80^{+0.04}_{-0.10}$~keV. The truncation of the inner disk and the decrease in $R_{in}$ with increasing luminosity is a prediction of the standard paradigm for the faint hard state that a hot ADAF or other coronal flow appears when the inner edge of the disk recedes from the ISCO \citep{narayan1994advection, esin1997advection}. In our best fit, we observe that during the decay, values of $R_{in}$ are all between 3 and 15~$R_g$, with a tentative increase towards the end of the outburst. To test the statistical significance of this tentative variation, we perform another fit in which the inner radii except for the last one are tied together, and we find $R_{in,1-5}=1.6^{+0.4}_{-0.3}$~$R_g$, and $R_{in,6}=12.2^{+8.4}_{-7.7}$, with C-stat increasing by 8 and $\chi^2$ increasing by 21 for 4 extra d.o.f. This test suggests that the crucial value of $R_{in,6}$ determining the evolution with regard to the luminosity is not statistically significant. We also find the spectrum
becomes harder with the photon index dropping from 1.72 to 1.62 when the
luminosity decreases, while the ionization parameter in \relxillCp\ is
reduced from $\xi\simeq2200$ to $\xi\simeq900$~ergs~cm~s$^{-1}$. 
%-+

%-+-------------------------------------------------------------------------------
%-+-------------------------------------------------------------------------------

%-+-------------------------------------------------------------------------------
%-+-------------------------------------------------------------------------------

%-+-------------------------------------------------------------------------------
\begin{table*}
\begin{center}
\caption{Best fit parameter values of model
\texttt{const*Tbabs*(diskbb+nthComp+relxillCp+xillverCp)} in a simultaneous fit
for the 2015 dataset (2015-M1). \label{tab:2015_M1_cal}}
{\renewcommand{\arraystretch}{1.2}%
\begin{tabular}{ccccccc}\hline \hline
Parameter&Obs.1&Obs.2&Obs.3&Obs.4&Obs.5&Obs.6\\
\hline
$N_H$ ($10^{21}$cm$^{-2}$)&\multicolumn{6}{c}{$4.12^{+0.08}_{-0.12}$}\\
$a_*$&\multicolumn{6}{c}{0.998}\\
$i$ (deg)&\multicolumn{6}{c}{$39.2^{+2.0}_{-1.8}$}\\
$A_{Fe}$&\multicolumn{6}{c}{$8.2\pm1.0$}\\
$C_{FPMA}$&\multicolumn{6}{c}{$1$}\\
$C_{FPMB}$&\multicolumn{6}{c}{$1.015\pm0.002$}\\
\hline
$\Gamma$&$1.724^{+0.011}_{-0.009}$&$1.667^{+0.012}_{-0.009}$&$1.628^{+0.013}_{-0.014}$&$1.646^{+0.008}_{-0.019}$&$1.605^{+0.009}_{-0.010}$&$1.624^{+0.007}_{-0.013}$\\
$kT_e$ (keV)&$>195$&$>224$&$>95$&$>67$&$>152$&$46^{+35}_{-8}$\\
$kT_{in} (keV)$&$0.46^{+0.03}_{-0.01}$&$0.30^{+0.03}_{-0.06}$&$0.45^{+0.02}_{-0.05}$&$0.36^{+0.08}_{-0.04}$&$0.058^{+0.023}_{-0.006}$&$0.80^{+0.04}_{-0.10}$\\
$R_{in}$($R_{ISCO}$)&$2.5\pm0.6$&$<2.3$&$<2.2$&$<2.4$&$3.5^{+1.4}_{-0.9}$&$12.4^{+8.4}_{-7.5}$\\
$log\xi$&$3.34^{+0.04}_{-0.02}$&$3.11^{+0.06}_{-0.05}$&$3.12^{+0.16}_{-0.07}$&$3.02^{+0.02}_{-0.27}$&$3.08^{+0.07}_{-0.06}$&$2.95^{+0.15}_{-0.49}$\\
$N_{disk}$&$170^{+41}_{-36}$&$58^{+32}_{-12}$&$30^{+12}_{-8}$&$<31$&$>3\times10^4$&$<1.1$\\
$N_{nthComp}$&$0.103^{+0.001}_{-0.003}$&$0.107^{+0.003}_{-0.006}$&$0.072^{+0.002}_{-0.003}$&$0.066^{+0.001}_{-0.007}$&$0.049\pm0.001$&$0.017^{+0.002}_{-0.001}$\\
$N_{relxillCp}$($10^{-3}$)&$1.15^{+0.16}_{-0.09}$&$1.16^{+0.24}_{-0.14}$&$0.84^{+0.18}_{-0.10}$&$0.60^{+0.25}_{-0.16}$&$0.32\pm0.07$&$0.08^{+0.03}_{-0.02}$\\
$N_{xillverCp}$($10^{-5}$)&$<12$&$<10$&$7.2^{+7.6}_{-7.0}$&$7.2^{+6.2}_{-5.1}$&$9.1^{+5.0}_{-4.6}$&$3.1^{+2.1}_{-2.2}$\\
$C_{XRT}$&$1.017\pm0.017$&$1.027\pm0.012$&$1.086\pm0.017$&$1.06\pm0.03$&$1.044\pm0.025$&$0.88\pm0.04$\\
\hline
$L/L_{edd}$ (\%)&2.0&1.8&1.7&1.2&1.0&0.5\\
$F_{disk}/F_{unabsorbed}$ (\%)&2.0&0.8&0.4&0.005&0&1.5\\
$R_s$&0.22&0.18&0.15&0.13&$0.10$&0.04\\
\hline
$C-stat$&\multicolumn{6}{c}{$10800$}\\
$\chi^2/d.o.f.$&\multicolumn{6}{c}{$12077/10730=1.126$}\\
\hline
\end{tabular}}
\end{center}
\raggedright{\textbf{Notes.} \\Luminosity calculated using unabsorbed flux
between $0.1-300$~keV, assuming a distance of 8 kpc and a black hole mass of 10~
$M_{\odot}$. The flux ratio of disk emission and the total unabsorbed one is
calculated in the $2-20$~keV range. The reflection strength $R_s$ is determined from the
flux ratio between \relxillCp\ and \nthComp\ in the energy range
of $20-40$~keV.}
\end{table*}
%-+-------------------------------------------------------------------------------

We also tried other emissivity profiles:
\begin{itemize}
\item Free emissivity index $q_1$ within the breaking radius $R_{br}$ free,
and a fixed outer emissivity index $q_2=3$. We find $q_1$ is between
the value of 3 and 4, $R_{br}$ could not be constrained and the other
parameters were insignificantly affected, with the C-stat decreasing by only
$\sim$23 for 12 fewer degrees of freedom. 
\item Free emissivity index $q_1=q_2$ all over the disk. We again find $q$
falls between 3 and 4, the other parameters were insignificantly affected, with
the C-stat decreasing by only $\sim$9 for 6 fewer degrees of freedom. 
\item Lamppost geometry. The fit is statistically worse by a
$\Delta$C-stat$=76$ for 6 fewer degrees of freedom. The corona height was found
to be fairly large ($10-20R_g$) and poorly constrained. 
\end{itemize}
%-+

\begin{figure}
\centering
\includegraphics[width=\linewidth]{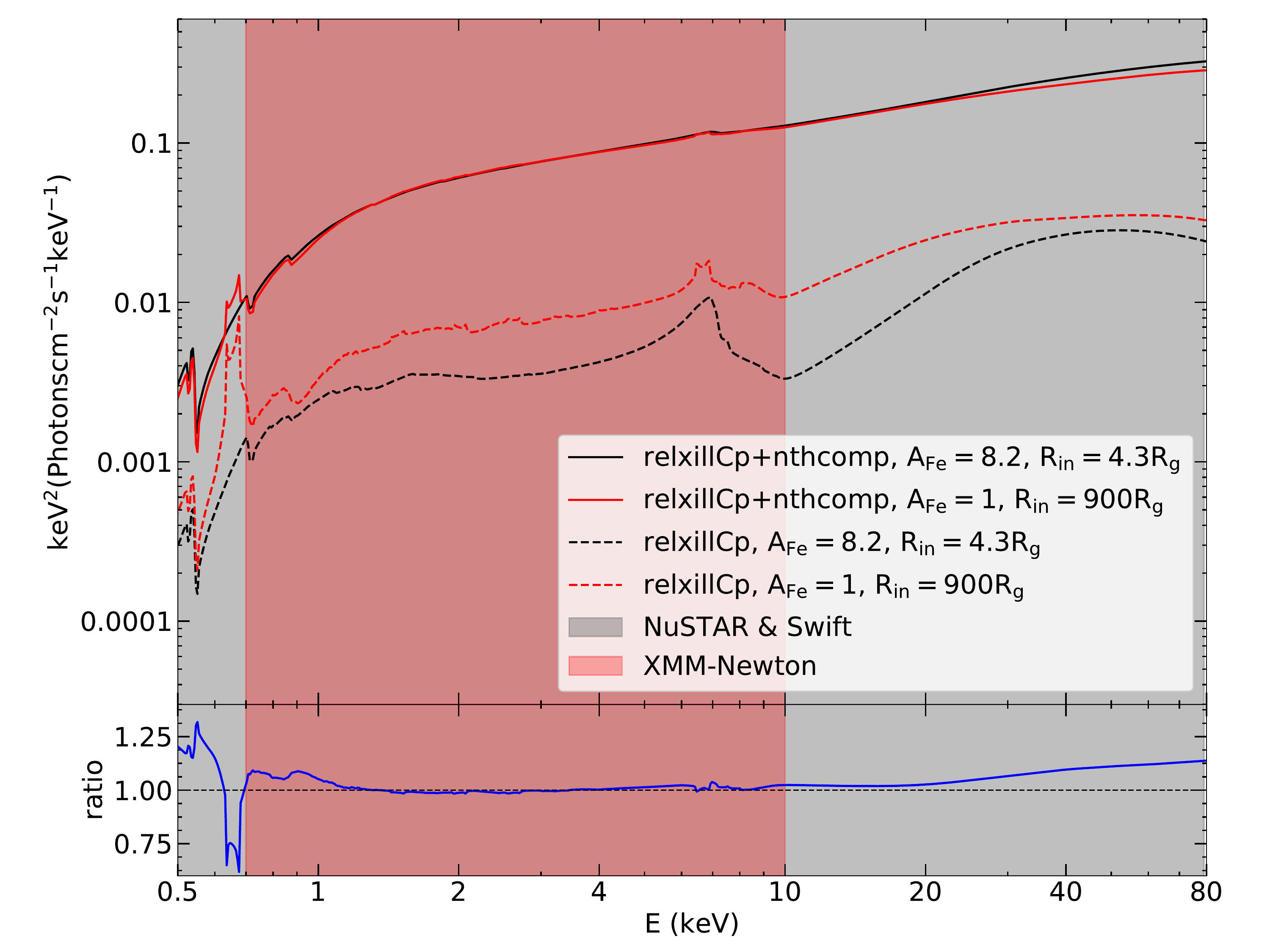}
\caption{Model components {\tt nthcomp+relxillCp} for the two cases: (1)
$A_{Fe}=8.2$ and $R_{in}=4.3^{+1.7}_{-1.1} R_{g}$; (2) $A_{Fe}=1.0$ and
$R_{in}=900 R_{g}$. The lower panel shows the ratio between the model component
{\tt nthcomp+relxillCp} in case (1) and case (2). It might be difficult to
distinguish these two cases when good quality data covering the oxygen emission
line below 0.7~keV and the Compton hump above 20~keV are not both available.
\label{fig:AFe}}
\end{figure}

We notice the large iron over-abundance in our fits: $8.2\pm1.0$ in Solar
units. To show the data prefers the over-abundance, we fix the iron abundance
to be the Solar value for this dataset (2015-M1-AFe1), and find the C-stat
increases by 791, for one additional degree of freedom. The disk becomes more
truncated, especially for Obs.5, in which the value of $R_{in}$ increases from
$4.3^{+1.7}_{-1.1} R_{g}$ to $>172 R_g$ (see Table~\ref{tab:2015_M1_AFe1_cal}
for the best-fit parameters). To interpret this, and following the procedure in
Section 6.1.4 in \citealt{javier_gx339}, we plot the model components {\tt
nthcomp+relxillCp} for these two cases in Figure~\ref{fig:AFe}, which shows it
could be difficult to distinguish a case with a Solar iron abundance and a disk
truncated at hundreds of $R_g$ from the case of iron over-abundance and mild
truncation, without good quality data covering the oxygen emission line below
0.7~keV and the Compton hump above 20~keV. Because of the low S/N of the
\swift\ data, we can not probe the oxygen line. However, with \nustar's wide
energy coverage up to 79~keV, we can see evidence of discrepancy above $\sim
30$~keV when the iron abundance is fixed at the Solar value, as shown in Figure
\ref{fig:M1_cal_r}. This demonstrates the preference of these data to require
large iron abundance.
%-+

%-+-------------------------------------------------------------------------------
%-+-------------------------------------------------------------------------------

\subsubsection{Model 2: taking the Comptonization of reflection into account \label{sec:M2}}

The presence of a corona as the source of the hard photons in the continuum
suggests the possibility for some of the reflected photons to intercept such a
corona before they reach the observer. This will result in additional Compton
scattering of some fraction of the reflection spectrum. As a first-order
adjustment, we can convolve the reflection spectrum with a Compton-scattering
kernel. For this we use the model
\simplcut\footnote{\url{http://jfsteiner.synology.me/wordpress/simplcut/}},
which adopts a scattering kernel based upon \nthComp\
\citep{zdziarski1996broad}. It has four physical parameters: the scattered
fraction $f_{sc}$, the spectral index $\Gamma$, the electron temperature $kT_e$,
and the reflection fraction $R_f$. We follow the procedures in
\citet{steiner2017self}, but we do not implement any linking between the
\diskbb\ parameters in the hard and soft states. In {\sc xspec} notation,
the model we adopt is:
%-+ 

\texttt{constant*Tbabs*[simplcut*(diskbb+relxillCp)}\\ \texttt{+xillverCp]} (2015-M2).

Here, in applying \simplcut\ in this way we are assuming that the
fraction of disk photons that are up-scattered in the corona is the same as the
fraction of reflected photons also intercepted by the corona, as they are
governed by one single scattering fraction. The best-fit parameters are shown
in Table \ref{tab:2015_M2_mod_cal}. For the last observation with the lowest
luminosity, the fit is consistent with the whole range of inner radii,
$1.5-800$~$R_g$, at the 90\% confidence level. This might be due to that the
scattering fraction is so large ($>0.97$) that the reflection features
including the iron line are heavily diluted, while the unblurred reflection
component \xillverCp\ can compensate for the iron emission seen in the
spectrum with a small ionization parameter ($log\xi<2.36$). The iron line
profile becomes difficult to determine and thus, the inner edge of the disk is
unconstrained. Also, the disk component is not evident in data.
%-+

In this framework of coronal Comptonization, there are several model
components: the power-law continuum, the intrinsic disk emission, the
relativistic reflection, and the Comptonized reflection. Besides the overall
normalization, only two parameters determine the relative strength of each
component: the scattered fraction $f_{sc}$, and the reflection fraction $R_f$.
The former depends on the geometry of the disk-corona system and also the
optical depth in the corona; while the latter is only associated with the
geometry of the system. We find that $f_{sc}$ increases when the luminosity
decreases (see Table~\ref{tab:2015_M2_mod_cal}). This could be explained by
changes in the corona structure. Figure~\ref{fig:2015_eem} ({\it right}) shows
how the model components change through observations. We calculated the
reflection strength as defined in \citealt{dauser2016normalizing}, and find
that except for observation~1, the other five observations show a decreasing
trend from $\sim4$ to $\sim0.2$, which is in line with the increasing inner
radius of the accretion disk.
%-+

%-+-------------------------------------------------------------------------------
\begin{table*}[htb!]
\begin{center}
\caption{Best fit parameter values of model
{\tt const*Tbabs*[simplcut*(diskbb+relxillCp)+xillverCp]} in the
simultaneous fit performed on the 2015 outburst dataset (2015-M2).
\label{tab:2015_M2_mod_cal} }
{\renewcommand{\arraystretch}{1.2}%
\begin{tabular}{ccccccc}\hline \hline
Parameter&Obs.1&Obs.2&Obs.3&Obs.4&Obs.5&Obs.6\\
\hline
$N_H$ ($10^{21}$cm$^{-2}$)&\multicolumn{6}{c}{$4.43^{+0.12}_{-0.06}$}\\
$a_*$&\multicolumn{6}{c}{0.998}\\
$i$ (deg)&\multicolumn{6}{c}{$39.2^{+1.6}_{-1.5}$}\\
$A_{Fe}$&\multicolumn{6}{c}{$7.7^{+1.0}_{-0.9}$}\\
$C_{FPMA}$&\multicolumn{6}{c}{1}\\
$C_{FPMB}$&\multicolumn{6}{c}{$1.0148\pm0.0018$}\\
\hline
$\Gamma$&$1.781^{+0.009}_{-0.008}$&$1.717\pm0.008$&$1.663\pm0.007$&$1.663^{+0.040}_{-0.007}$&$1.635^{+0.004}_{-0.006}$&$1.654^{+0.023}_{-0.032}$\\
$f_{sc}$&$0.51\pm0.02$&$0.64\pm0.02$&$0.68^{+0.04}_{-0.03}$&$0.66^{+0.06}_{-0.08}$&$0.45^{+0.08}_{-0.05}$&$>0.97$\\
$kT_e$ (keV)&$>196$&$>100$&$>66$&$67^{+47}_{-18}$&$>117$&$46^{+111}_{-12}$\\
$kT_{in} (keV)$&$0.51\pm0.03$&$0.66^{+0.11}_{-0.12}$&$>0.57$&$<0.13$&$0.110^{+0.018}_{-0.002}$&$>0.78$\\
$N_{disk}$&$227^{+74}_{-26}$&$26^{+55}_{-8}$&$8^{+11}_{-2}$&$(4.8^{+1.8}_{-4.5})\times10^4$&$>7.1\times10^4$&$20^{+8}_{-1}$\\
$R_{in}$($R_{ISCO}$)&$<1.9$&$1.8^{+3.0}_{-0.6}$&$<1.9$&$<2.1$&$5.0^{+2.7}_{-1.4}$&$-$\\
$log\xi$&$3.29^{+0.04}_{-0.06}$&$3.07^{+0.07}_{-0.05}$&$3.17^{+0.18}_{-0.07}$&$3.04\pm0.05$&$2.42^{+0.40}_{-0.29}$&$<2.36$\\
$N_{relxillCp}$($10^{-3}$)&$2.7^{+0.2}_{-0.4}$&$2.8^{+0.3}_{-0.5}$&$2.0\pm0.4$&$1.2^{+0.2}_{-0.5}$&$0.59^{+0.15}_{-0.11}$&$<0.64$\\
$N_{xillverCp}$($10^{-5}$)&$<8.3$&$<7.3$&$6.5^{+4.2}_{-5.1}$&$8.0^{+4.4}_{-4.3}$&$8.5^{+4.4}_{-4.3}$&$6.8\pm1.9$\\
$C_{XRT}$&$1.018^{+0.016}_{-0.015}$&$1.007\pm0.016$&$1.091^{+0.013}_{-0.007}$&$1.038\pm0.025$&$1.042^{+0.028}_{-0.025}$&$0.94\pm0.04$\\
\hline
$L/L_{edd}$ (\%)&2.0&1.8&1.7&1.2&1.0&0.5\\
$R_s$&1.87&4.16&4.13&0.83&$0.21$&0.18\\
\hline
$C-stat$&\multicolumn{6}{c}{$10822$}\\
$\chi^2/d.o.f.$&\multicolumn{6}{c}{$12067/10730=1.125$}\\
\hline
\end{tabular}}
\end{center}
\raggedright{\textbf{Notes.} \\Luminosity calculated using unabsorbed flux
between $0.1-300$~keV, assuming a distance of 8~kpc and a black hole mass of
10~$M_{\odot}$. The reflection strength $R_s$ is determined from the flux ratio
between \relxillCp\ and \nthComp\ in the energy range of $20-40$~keV.}
\end{table*}
%-+-------------------------------------------------------------------------------

\subsection{The 2013 dataset: rise and decay in a failed outburst}

The absorbed power-law fit on the 2013 data do not show any strong indication
of the existence of a soft disk component, thus we started the fit by fixing
the disk temperature to 0.05~keV.  However, with free disk temperatures the
fit goes down in C-stat by 725 with 10 less d.o.f., which is a significant
improvement. The flux ratio in the $2-20$~keV range between the intrinsic disk emission
and the unabsorbed total one is around 3\%, which matches the expected faint
disk in the low-hard state, but the determined disk temperatures are above
0.8~keV for the last three observations. In addition, the inner edge of the
disk does not follow a one-way trend with luminosity. The best-fit parameters
for this model (2013-M1) are shown in Table~\ref{tab:2013_M1_cal}.
%-+

\begin{figure*}
\epsscale{1.15}
\plottwo{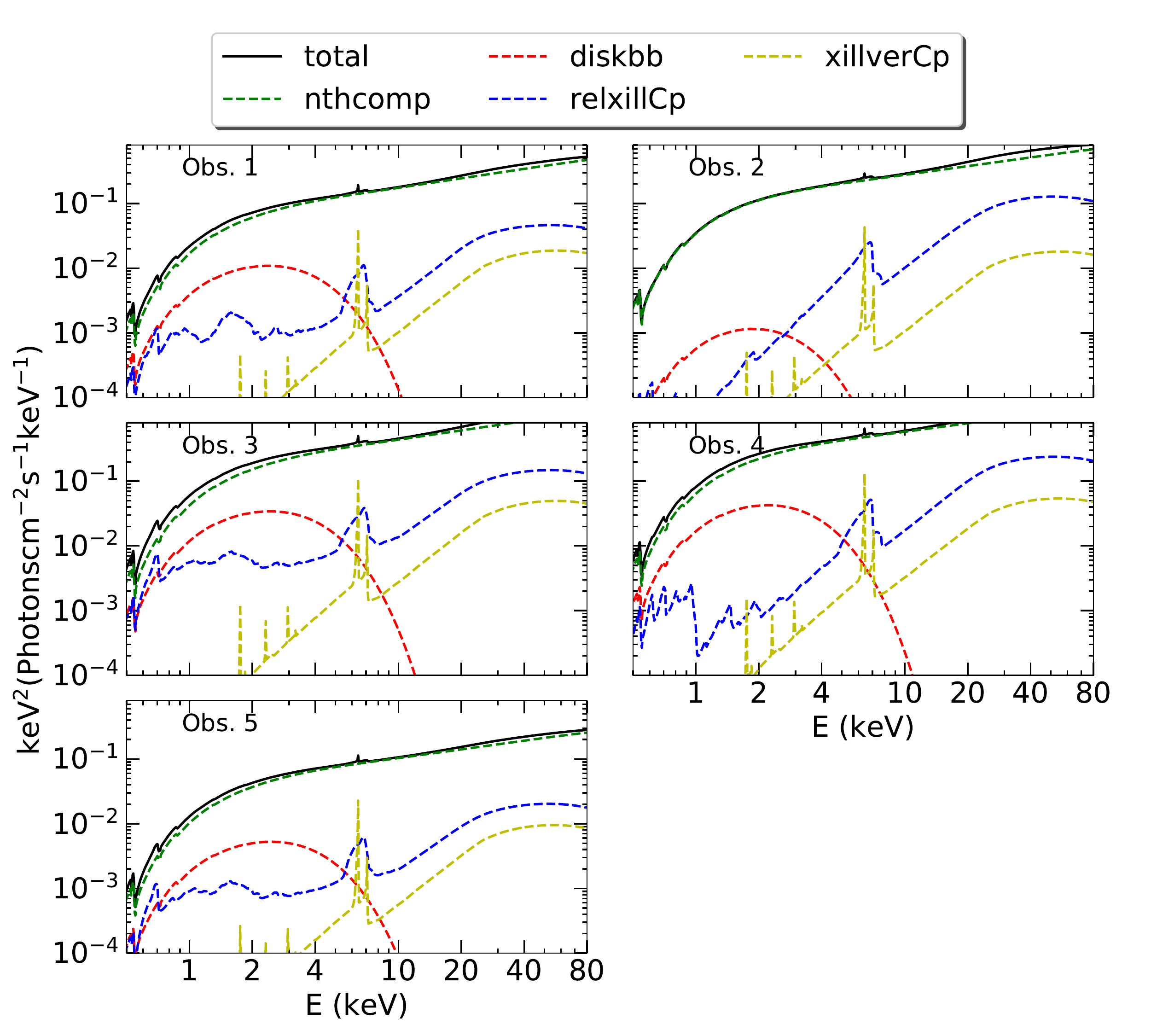}{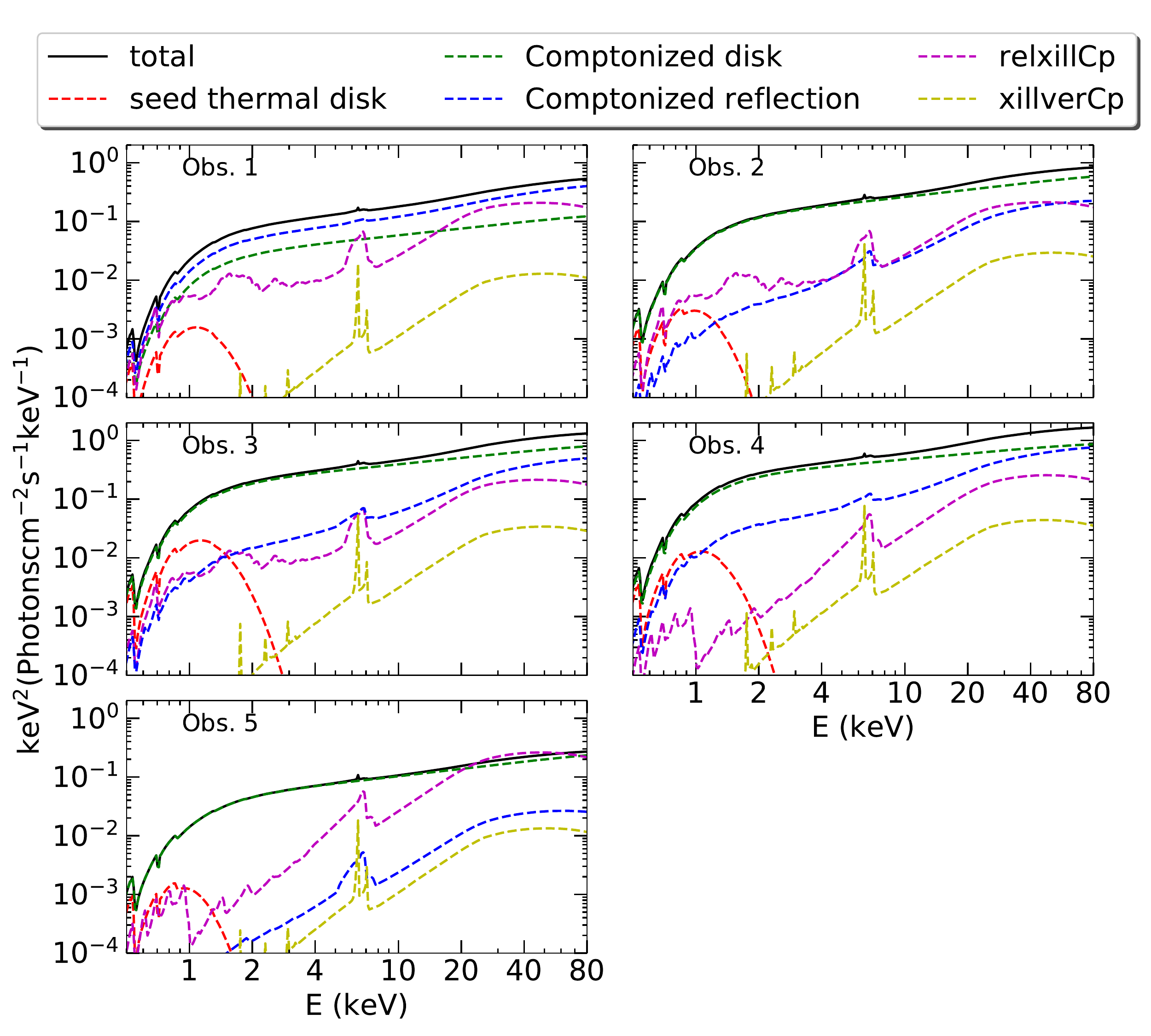}
%\caption{Model components for individual observation in 2013 for M1 ({\it left}) and M2 ({\it right}). ({\it left}) Black: total, red: \diskbb, green: \nthComp, blue: \relxillCp, yellow: \xillverCp. ({\it right}) Black: total, red: the intrinsic disk emission (\diskbb), magenta: the reflection component (\relxillCp), green: Comptonized disk including the unscattered intrinsic disk emission and the continuum component it generates (\simplcut\ convolved with \diskbb), blue: Comptonized reflection including the unscattered reflection and the Comptonized component it generates (\simplcut\ convolved with \relxillCp), yellow: \xillverCp.}
\caption{Model components for individual observation in 2013 for M1 ({\it left}) and M2 ({\it right}). The component each color represents is indicated in the figure. In M1, the determined disk temperatures are above 0.8~keV for the last three observations, while in M2, the disk temperatures fall into a range of values closer to the expectation for this source ($kT_{in}\lesssim0.2$ keV), and the intrinsic disk flux becomes much smaller which is more typical for the low-hard state. }
\label{fig:2013_eem}
\end{figure*}

We then try the model taking the Comptonization of reflection into account in
this dataset (2013-M2), following the same procedures as in Section~\ref{sec:M2}.
Compared to 2013-M1, C-stat increases by 224 with the same d.o.f.
which is statistically worse; but we also notice that M2 reduced $\chi^2$ by 7.
Additionally, this model provides a more reasonable combination of disk and
power-law components. As shown in Table~\ref{tab:2013_M2}, the disk temperatures
fall into a range of values closer to the expectation for this source
($kT_{in}\lesssim0.2$ keV). In Figure~\ref{fig:2013_eem} ({\it right}), the
intrinsic disk flux becomes much smaller which is more typical for the low-hard
state.
%-+

%-+-------------------------------------------------------------------------------
%-+-------------------------------------------------------------------------------

%-+-------------------------------------------------------------------------------
\begin{table*}
\begin{center}
\caption{Best fit parameter values of model
{\tt const*Tbabs*(diskbb+nthComp+relxillCp+xillverCp)} in a simultaneous fit
performed on the 2013 outburst dataset (2013-M1). \label{tab:2013_M1_cal}}
{\renewcommand{\arraystretch}{1.2}%
\begin{tabular}{cccccc}\hline \hline
Parameter&Obs.1&Obs.2&Obs.3&Obs.4&Obs.5\\
\hline
$N_H$ ($10^{21}$cm$^{-2}$)&\multicolumn{5}{c}{$4.12^{+0.06}_{-0.18}$}\\
$a_*$&\multicolumn{5}{c}{0.998}\\
$i$ (deg)&\multicolumn{5}{c}{$40.7^{+0.7}_{-0.8}$}\\
$A_{Fe}$&\multicolumn{5}{c}{$3.83\pm0.06$}\\
$C_{FPMA}$&\multicolumn{5}{c}{1}\\
$C_{FPMB}$&\multicolumn{5}{c}{$1.0219\pm0.0009$}\\
\hline
$\Gamma$&$1.56\pm0.02$&$1.585\pm0.001$&$1.606\pm0.001$&$1.54\pm0.02$&$1.616\pm0.001$\\
$kT_e$ (keV)&$>473$&$231^{+38}_{-21}$&$>540$&$>620$&$>497$\\
$kT_{in} (keV)$&$0.422\pm0.002$&$0.53^{+0.08}_{-0.02}$&$0.892\pm0.002$&$0.796\pm0.001$&$0.80\pm0.17$\\
$R_{in}$($R_{ISCO}$)&$<1.5$&$3.9\pm0.8$&$14.0^{+3.5}_{-3.1}$&$10.0^{+1.6}_{-1.5}$&$32.3^{+17.2}_{-10.9}$\\
$log\xi$&$0.70^{+0.07}_{-0.06}$&$1.01^{+0.03}_{-0.06}$&$1.69^{+0.03}_{-0.49}$&$1.54^{+0.04}_{-0.13}$&$2.97^{+0.04}_{-0.08}$\\
$N_{disk}$&$3.4\pm0.2$&$-$&$7.4\pm0.3$&$16.8\pm0.5$&$1.05\pm0.05$\\
$N_{nthComp}$&$0.054\pm0.019$&$0.071^{+0.014}_{-0.012}$&$0.0920\pm0.0001$&$0.1351\pm0.0001$&$0.02039\pm0.00003$\\
$N_{relxillCp}$($10^{-3}$)&$0.63\pm0.03$&$1.5\pm0.2$&$2.0\pm0.4$&$2.7\pm0.6$&$0.23\pm0.03$\\
$N_{xillverCp}$($10^{-4}$)&$2.7\pm0.3$&$2.8\pm0.4$&$5.0\pm0.5$&$6.6\pm0.4$&$1.1\pm0.1$\\
$C_{XRT}$&$1.057\pm0.025$&$1.174^{+0.015}_{-0.014}$&$0.982\pm0.015$&$1.039\pm0.010$&$1.074\pm0.010$\\
\hline
$L/L_{edd}$ (\%)&1.4&2.4&3.6&4.6&0.8\\
$F_{disk}/F_{unabsorbed}$ (\%)&2.7&0.1&3.4&2.8&2.3\\
$R_s$&0.12&0.21&0.15&0.19&$0.09$\\
\hline
$C-stat$&\multicolumn{5}{c}{$9556$}\\
$\chi^2/d.o.f.$&\multicolumn{5}{c}{$10253/9336=1.098$}\\
\hline
\end{tabular}}
\end{center}
\vspace{0.3cm}\textbf{Notes.} \\Luminosity calculated using unabsorbed flux
between $0.1-300$~keV, assuming a distance of 8 kpc and a black hole mass of 10~
$M_{\odot}$. The flux ratio of disk emission and the total unabsorbed one is
calculated in the $2-20$~keV range. The reflection strength $R_s$ is determined from the
flux ratio between \relxillCp\ and \nthComp\ in the energy range
of $20-40$~keV.
\end{table*}
%-+-------------------------------------------------------------------------------

%-+-------------------------------------------------------------------------------
\begin{table*}[h!]
\begin{center}
\caption{Best fit parameter values of model
{\tt const*Tbabs*[simplcut*(diskbb+relxillCp)+xillverCp]} in a simultaneous
fit performed on the 2013 outburst dataset (2013-M2). \label{tab:2013_M2}}
{\renewcommand{\arraystretch}{1.2}%
\begin{tabular}{cccccc}\hline \hline
Parameter&Obs.1&Obs.2&Obs.3&Obs.4&Obs.5\\
\hline
$N_H$ ($10^{21}$cm$^{-2}$)&\multicolumn{5}{c}{$6.85^{+0.10}_{-0.09}$}\\
$a_*$&\multicolumn{5}{c}{0.998}\\
$i$ (deg)&\multicolumn{5}{c}{$39.7^{+2.6}_{-2.0}$}\\
$A_{Fe}$&\multicolumn{5}{c}{$2.82^{+0.17}_{-0.15}$}\\
$C_{FPMA}$&\multicolumn{5}{c}{1}\\
$C_{FPMB}$&\multicolumn{5}{c}{$1.0219\pm0.0012$}\\
\hline
$\Gamma$&$1.640^{+0.011}_{-0.010}$&$1.635^{+0.010}_{-0.009}$&$1.676\pm0.006$&$1.705\pm0.007$&$1.626^{+0.005}_{-0.004}$\\
$f_{sc}$&$0.78^{+0.06}_{-0.07}$&$0.69\pm0.06$&$0.79\pm0.03$&$0.78^{+0.02}_{-0.03}$&$0.31^{+0.05}_{-0.01}$\\
$kT_e$ (keV)&$>148$&$>159$&$>272$&$>142$&$>210$\\
$kT_{in} (keV)$&$0.130^{+0.011}_{-0.024}$&$0.130^{+0.006}_{-0.009}$&$0.204^{+0.008}_{-0.020}$&$0.156^{+0.006}_{-0.028}$&$0.116\pm0.002$\\
$N_{disk}$($10^4$)&$<2.0$&$8.8^{+0.3}_{-1.0}$&$1.6^{+3.2}_{-1.3}$&$8.1^{+0.6}_{-0.7}$&$>9.4$\\
$R_{in}$($R_{ISCO}$)&$>11.4$&$4.4^{+1.7}_{-1.0}$&$14.3^{+7.4}_{-6.1}$&$12.6^{+4.7}_{-3.6}$&$15.6^{+14.7}_{-5.9}$\\
$log\xi$&$2.69\pm0.02$&$<1.81$&$1.76^{+0.25}_{-0.34}$&$2.00^{+0.02}_{-0.12}$&$<1.78$\\
$N_{relxillCp}$($10^{-3}$)&$2.5^{+0.2}_{-0.9}$&$3.0^{+0.3}_{-0.6}$&$6.1^{+0.7}_{-0.6}$&$9.2^{+0.4}_{-0.5}$&$0.33^{+0.27}_{-0.09}$\\
$N_{xillverCp}$($10^{-4}$)&$<2.2$&$3.0^{+1.3}_{-1.6}$&$3.9^{+1.7}_{-1.8}$&$5.3^{+1.4}_{-1.5}$&$1.5\pm0.5$\\
$C_{XRT}$&$1.025\pm0.024$&$1.140\pm0.015$&$0.947^{+0.016}_{-0.015}$&$1.012^{+0.012}_{-0.011}$&$1.028^{+0.011}_{-0.012}$\\
\hline
$L/L_{edd}$ (\%)&1.4&2.4&3.6&4.6&0.8\\
$R_s$&1.99&0.50&0.74&0.99&$0.14$\\
\hline
$C-stat$&\multicolumn{5}{c}{$9780$}\\
$\chi^2/d.o.f.$&\multicolumn{5}{c}{$10246/9336=1.097$}\\
\hline
\end{tabular}}
\end{center}
\raggedright{\textbf{Notes.} \\Luminosity calculated using unabsorbed flux
between $0.1-300$~keV, assuming a distance of 8 kpc and a black hole mass of 10~
$M_{\odot}$. The reflection strength $R_s$ is determined from the
flux ratio between \relxillCp\ and \nthComp\ in the energy range
of $20-40$~keV.}
\end{table*}
%-+-------------------------------------------------------------------------------

%-+-------------------------------------------------------------------------------
\begin{table*}[htb!]
\caption{The intrinsic parameters of the system found in different simultaneous
fits in this paper: hydrogen column density ($N_H$), the dimensionless spin
parameter $a_*=0.998$ which is frozen in all through, the inclination of the
inner disk $i$, the iron abundance with respect to the solar value $A_{Fe}$.
The model description, C-stat and $\chi^2$ values are also
provided.\label{tab:global_par}}
{\renewcommand{\arraystretch}{1.2}
\begin{tabular}{ccccccc}\hline \hline
Fit&Model Desciption&C-stat&$\chi^2$/d.o.f.&$N_H$&$i$&$A_{Fe}$\\
&&&&($10^{21}$cm$^{-2}$)&(deg)&\\
\hline
\hline
2015-M1&Standard reflection model&10800&12077/10730&$4.12^{+0.08}_{-0.12}$&$39.2^{+2.0}_{-1.8}$&$8.2\pm1.0$\\
&(\texttt{diskbb+nthcomp+relxillCp+xillverCp})&&=1.126&&&\\
2015-M1-AFe1&Standard reflection model,&11591&12584/10731&$4.53^{+0.04}_{-0.05}$&$75\pm5$&$1.0$\\
&$A_{Fe}=1.0$&&=1.173&&&\\
2015-M2&Model considering the coronal Comptonization&10822&12067/10730&$4.43^{+0.12}_{-0.06}$&$39.7^{+2.6}_{-2.0}$&$7.7^{+1.0}_{-0.9}$\\
&[\texttt{simplcut*(diskbb+relxillCp)+xillverCp}]&&1.125&&&\\
\hline
2013-M1&Standard reflection model&9556&10253/9336&$4.12^{+0.06}_{-0.18}$&$40.7^{+0.7}_{-0.8}$&$3.83\pm0.06$\\
&&&=1.098&&&\\
2013-M2&Model considering the coronal Comptonization&9780&10246/9336&$6.85^{+0.10}_{-0.09}$&$39.7^{+2.6}_{-2.0}$&$2.82^{+0.17}_{-0.15}$\\
&&&=1.097&&&\\
\hline
\end{tabular}}
\end{table*}
%-+-------------------------------------------------------------------------------

\section{Discussion}\label{dis}

The parameters that are global to all observations are: the Galactic hydrogen
column density $N_H$, the spin parameter $a_*$, the inclination angle $i$ and
the iron abundance $A_{Fe}$. Table~\ref{tab:global_par} shows a summary of
these intrinsic parameter values found in different simultaneous fits performed
in this paper. The inclination is consistent with $i=40\pm2$ deg through all
fits except for 2015-M1-AFe1. Assuming that the inclination of the inner disk
is equal to the binary orbit inclination, with the latest measurement of the
mass function $f(M)=\frac{M_{bh}\sin^3i}{(1+q)^2}=1.91\pm0.08$~$M_\odot$ and
$q=\frac{M_c}{M_{bh}}=0.18\pm0.05$ \citep{heida2017potential}, we estimate the mass of the black
hole to be $M_{bh}=10.0\pm0.6$~$M_\odot$.
%-+

\citet{furst2015complex} found the disk to be truncated at tens of $R_g$, based
on the same \nustar\ and \swift\ dataset of the 2013 outburst, using the model 
{\tt constant*tbabs*[powerlaw+relconv(reflionx)+}\\ {\tt gaussian]}, which includes the older reflection model {\tt reflionx}
\citep{ross2005comprehensive}, convolved with the relativistic kernel {\tt
relconv} \citep{dauser2010broad}. 
By comparing the {\tt simplcut*relxillCp} with the \relxillCp\ models shown in
Figures~\ref{fig:2015_eem} and \ref{fig:2013_eem} (\textit{right}), the slope of the reflection
component is reduced as a pure consequence of coronal scattering. This could
potentially explain the results found in \citealt{furst2015complex}.
After allowing a difference between the photon index feeding the reflection
($\sim1.3$) and the one in the power-law continuum ($\sim1.6$) to account for a possible physically extended corona with a non-uniform temperature profile, they found the iron abundance was
also reduced (from $\sim5$ to $\sim1.5$), and thus, forces the disk to be much more
truncated to minimize the relativistic effects that blur the line profile. 
Nevertheless, in our case, M2 only provides a significant
reduction in the iron abundance compared to M1 for the 2013 data.

We do not observe a clear evolution for a decrease of disk temperature with decreasing luminosity, as another prediction in the truncation disk scenario. The reasons for this are threefold. First, the \textit{Swift} data have the total numbers of counts much smaller than \nustar\ ($10-100$ times smaller), which makes the determination of disk temperatures governed by the low energy range very difficult. Second, the large disk temperatures we find with M1 could be artificially produced by the
complexity of the Comptonization model \citep{kolehmainen2013soft}. 
In the frequency resolved spectra, the most rapidly variable part of the flow
has harder spectra and less reflection than the slowly variable emission \citep{axelsson2013imperfect}. 
This feature would give rise to spectral curvature in broad-band data (as seen in, e.g., \citealt{makishima2008suzaku}),
and thus, requires an additional soft component when such a continuum is fitted with a single Comptonization component. Lastly, as we do not observe a strong evolution pattern of the disk's inner radius with luminosity, it is understandable that the disk temperature does not evolve as expected either.
%+

%-+

%-+-------------------------------------------------------------------------------
\begin{figure*}
\centering
\includegraphics[width=\linewidth]{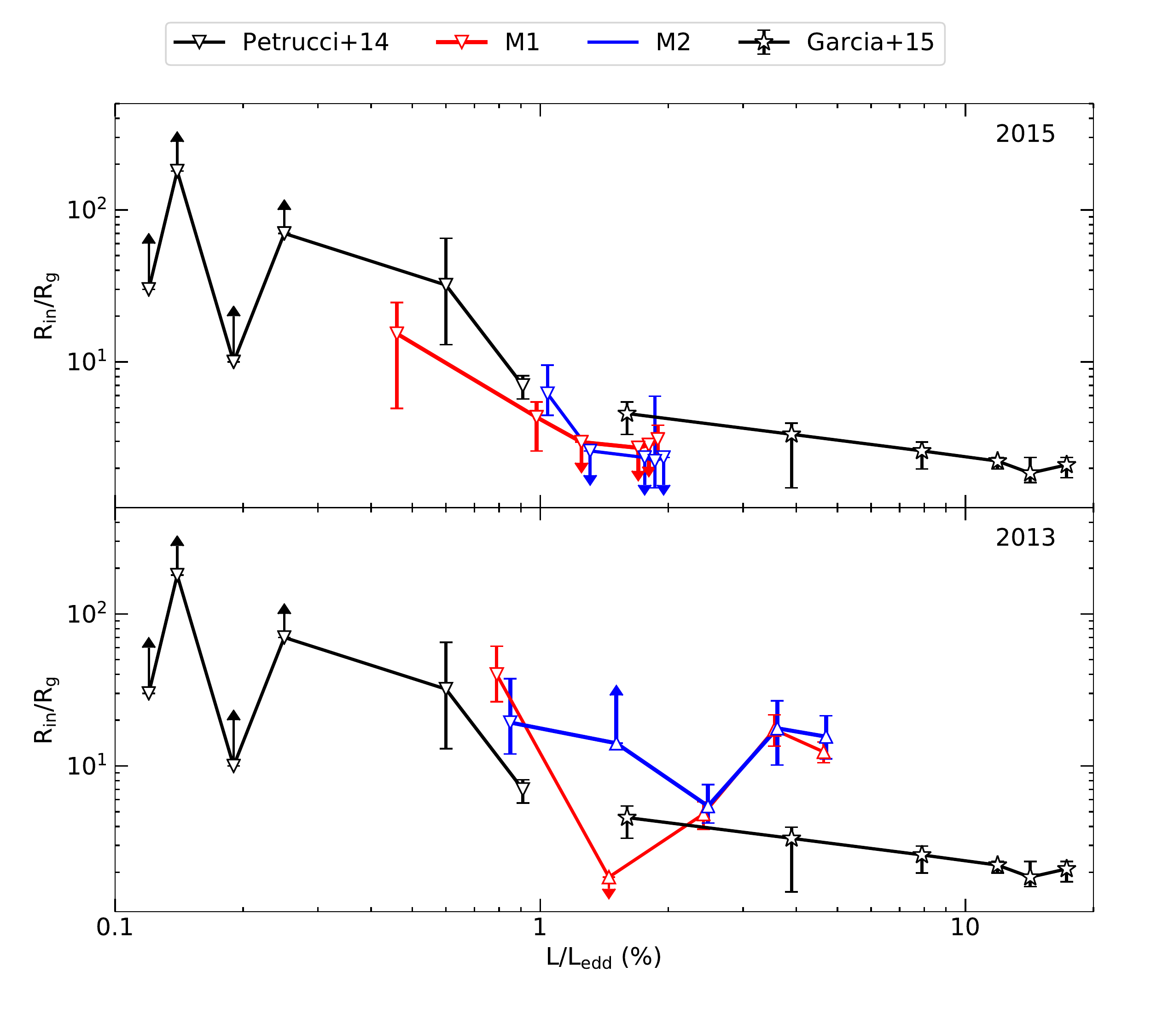}
\caption{Comparison for \gx339\ of our estimates (\textit{upper}: the 2015
dataset, \textit{lower}: the 2013 dataset) with those in the previous
literature \citet{javier_gx339} and \citet{petrucci2014return} of the
inner-disk radius vs. Eddington-scaled luminosity. The luminosity values for
the same observations are slightly shifted for clarity. \label{fig:Rin_ours}}
\end{figure*}
%-+-------------------------------------------------------------------------------

%-+-------------------------------------------------------------------------------
\begin{figure*}
\begin{center}
\includegraphics[width=1.0\textwidth]{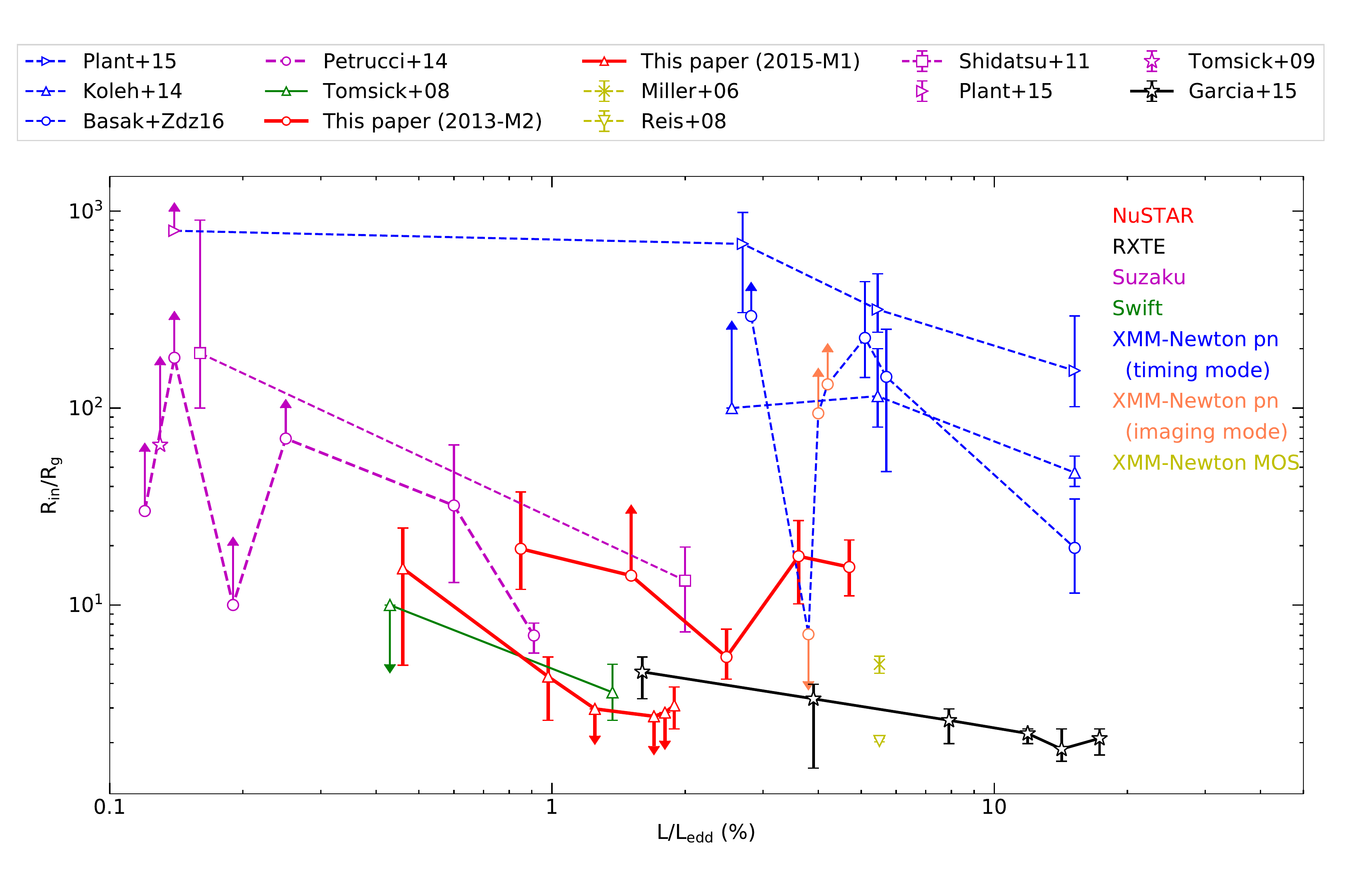}
\end{center}
\caption{Comparison of the inner-disk radius vs. Eddington-scaled luminosity
for \gx339. Our best fit values are shown in contrast with previous studies
using reflection spectroscopy
\citep{plant2015truncated,kolehmainen2013soft,basak2016spectral,petrucci2014return,tomsick2009truncation,
miller2006long, reis2008systematic, shidatsu2011x, tomsick2008broadband,
javier_gx339}.  Each instrument is plotted with a different color as indicated.
The luminosity values for the same observations are slightly shifted for
clarity.\label{fig:Rin_all}}
\end{figure*}
%-+-------------------------------------------------------------------------------

The evolution of the inner disk radius changing with respect to the luminosity
we find in different models, and those reported by \citet{javier_gx339} and
\citet{petrucci2014return}  are shown in Figure \ref{fig:Rin_ours}. For a
detailed summary of estimations of $R_{in}$ in previous literature for \gx339\
between a luminosity range of $0.1\%- 20\% L_{edd}$ in low-hard state obtained
from the reflection spectroscopy, see Table~5 in \citet{javier_gx339}.
%-+

Among all the fits we performed, 2015-M1 shows the most promising decreasing trend of $R_{in}$ with increasing luminosity. However, this result is not statistically significant, as we suggested in Section 3.1.1.
By comparing the trends M1 and M2 give for
the 2015 dataset (see the upper panel in Figure \ref{fig:Rin_ours}), except for
the one missing data point in M2 where $R_{in}$ is unconstrained, the other
five values agree well with each other, suggesting a consistent and
model-robust conclusion.
%-+

Another interesting aspect to notice is that in the luminosity range covered by
the two datasets, the values of $R_{in}$ found for the 2013 observations is
slightly larger. This could be due to the fact that the 2013 observations were
taken in the rising phase (obs.1-4), and at the end of a failed outburst
(obs.5), while the 2015 data was taken during the decay of a successful one.
The hysterisis pattern typically observed in the hardness-intensity diagram of
this source suggests that the evolution during the rising and decay phases
displays a different phenomenology, which is likely to affect the evolution of
the inner radius.
%-+

The evolution of $R_{in}$ with luminosity in the low-hard state is a matter of central importance for the study of black hole binaries. As our results are limited by the relatively small luminosity range we explore, we plot the reported results in previous literatures and our preferred ones (2015-M1 and 2013-M2) of inner radius vs. Eddington-scaled luminosity in Figure~\ref{fig:Rin_all}, sorted and colored
with regard to satellites, instruments, and observation mode.  At luminosities
larger than 1\% $L_{edd}$, there are two groups of results: an upper group with
inner radii between $20R_g$ and $800R_g$ comprised by values from \xmm\ pn
timing mode and two imaging mode data; and a bottom group with $R_{in}<20R_g$
aligned with \nustar\,, \rxte\,, \suzaku\,, \swift\,, \xmm\ MOS and one \xmm\ pn
imaging mode data. These results indicate the possibility of calibration issues
with \xmm\ pn timing mode data as the main factor responsible for the very
extreme truncation.
%-+

\section{Conclusions}\label{conclusion}

We have analysed eleven observations of \gx339\ in the low-hard state seen by
\nustar\ and \swift\,, five taken in a failed outburst in 2013 and the other six
during the decay of the 2015 outburst. The luminosity covers the range of 0.5\% to 5\% $L_{edd}$, which only covers a fraction of the usual luminosity range typically observed during the outburst for this source (up to $20-30\%$~$L_{edd}$). Each spectrum spans the energy range 3--79~keV
from \nustar, and 0.5--8~keV from \swift. The data have in total 10.7 million
counts, and a composed exposure time of 790~ks.
%-+

Both datasets are fitted with two models: a standard reflection model including
intrinsic disk emission, power-law continuum, and both the relativistic and
unblurred reflection components {\tt
const*Tbabs*(diskbb+nthComp+relxillCp+xillverCp)} (M1); and a model in which
the reflection component is Comptonized by the corona {\tt
constant*Tbabs*[simplcut*(diskbb+relxillCp)}\\ \texttt{+xillverCp]}.  

During the decay in 2015, with fit M1 we find that the inner disk recedes from the ISCO, values of $R_{in}$ are all between 3 and 15~$R_g$, with a tentative increase towards the end of the outburst, although we do notice that the largest truncation radius here is not statistical significant. Fit M2 provides similar results, except for the
last observation whose inner radius is unconstrained. As for the 2013 dataset,
the disk temperatures determined from M1 are unphysically large for these
luminosities in the low-hard state, while M2 can effectively reconcile these
values ($kT_{in}\lesssim0.2$~keV) and provide more physical trends. 
The evolution of $R_{in}$ with luminosity for
the 2013 data is somewhat less monotonic than for the 2015, and while the inner
radius is larger in the former, we find the largest disk truncation is
constrained to be less than $37R_g$ when the source is at 0.8\% $L_{edd}$.
%-+

\bigskip
Part of this work was carried out by JW during attendance to the Summer Undergraduate Research Fellowship (SURF) at California Institute of Technology in 2017. Warm hospitality and tutelage are kindly acknowledged, in particular from the members of the \nustar\ Science Operations Team (K. Foster, B. Grenfestette, K. Madsen, M. Heida, M. Brightman, and D. Stern). We would like to thank the referee for useful comments towards the improvement of this paper. 
We also thank B. de Marco and G. Ponti for useful discussions. 
JAG acknowledges support from NASA grants NNX17AJ65G and 80NSSC177K0515, and from the Alexander von Humboldt Foundation. JFS has been supported by NASA Hubble Fellowship grant
HST-HF-51315.01. SC is supported by the SERB National Postdoctoral Fellowship (No. PDF/2017/000841). We thank the \nustar\ Operations, Software, and Calibration teams for support with the execution and analysis of these observations.

\appendix

Table~\ref{tab:2015_M1_AFe1_cal} shows the best-fit parameters when we fix the iron abundance
to be the Solar value for the 2015 observations (2015-M1-AFe1). The disk becomes more truncated, especially for Obs.5, in which the value of $R_{in}$ increases from $4.3^{+1.7}_{-1.1} R_{g}$ to $>172 R_g$. However, the fit is significantly worse in statistics with regard to 2015-M1, with C-stat increasing by 791 for one additional degree of freedom. In addition, with \nustar's wide energy coverage up to 79~keV, we can see evidence of discrepancy above $\sim30$~keV, as shown in Figure~\ref{fig:M1_cal_r}. This demonstrates the preference of these data to require large iron abundance, and a systematic discussion about the iron over-abundance found by reflection spectroscopy will be presented in a future publication. 

\setcounter{table}{0}
\renewcommand{\thetable}{A\arabic{table}}
\begin{table*}
\begin{center}
\caption{Best fit parameter values of model
\texttt{const*Tbabs*(diskbb+nthComp+relxillCp+xillverCp)} with a frozen iron
abundance at the solar value in the simultaneous fit for the 2015 dataset
(2015-M1-AFe1). \label{tab:2015_M1_AFe1_cal}}
{\renewcommand{\arraystretch}{1.2}
\begin{tabular}{ccccccc}\hline \hline
Parameter&Obs.1&Obs.2&Obs.3&Obs.4&Obs.5&Obs.6\\
\hline
$N_H$ ($10^{21}$cm$^{-2}$)&\multicolumn{6}{c}{$4.53^{+0.04}_{-0.05}$}\\
$a_*$&\multicolumn{6}{c}{0.998}\\
$i$ (deg)&\multicolumn{6}{c}{$75.0\pm5.0$}\\
$A_{Fe}$&\multicolumn{6}{c}{$1.0$}\\
$C_{FPMA}$&\multicolumn{6}{c}{$1$}\\
$C_{FPMB}$&\multicolumn{6}{c}{$1.015^{+0.004}_{-0.002}$}\\
\hline
$\Gamma$&$1.767\pm0.002$&$1.70^{+0.05}_{-0.05}$&$1.665^{+0.006}_{-0.010}$&$1.665^{+0.007}_{-0.010}$&$1.637\pm0.001$&$1.653\pm0.002$\\
$kT_e$ (keV)&$>388$&$>381$&$>308$&$>250$&$>241$&$182^{+42}_{-31}$\\
$kT_{in} (keV)$&$<0.06$&$<0.13$&$0.31^{+0.08}_{-0.11}$&$0.34^{+0.23}_{-0.01}$&$0.059^{+0.003}_{-0.006}$&$0.752\pm0.003$\\
$N_{disk}$&$-$&$-$&$<86$&$<38$&$>3.2\times10^6$&$<1.08$\\
$R_{in}$($R_{g}$)&$<1.4$&$25.2^{+4.8}_{-8.8}$&$>18.7$&$36.1^{+9.5}_{-6.3}$&$>172$&$55.5^{+68.2}_{-17.6}$\\
$log\xi$&$3.321\pm0.001$&$3.22^{+0.05}_{-0.07}$&$3.20^{+0.10}_{-0.07}$&$3.002^{+0.006}_{-0.053}$&$3.027\pm0.006$&$2.75^{+0.02}_{-0.08}$\\
$N_{nthComp}$($10^{-3}$)&$75.2\pm0.3$&$108.8^{+0.2}_{-6.4}$&$77^{+5}_{-9}$&$65.7^{+0.2}_{-14.1}$&$49.1\pm0.1$&$17.4\pm0.1$\\
$N_{relxillCp}$($10^{-3}$)&$3.509^{+0.013}_{-0.009}$&$1.9^{+1.9}_{-0.9}$&$1.9^{+0.6}_{-0.2}$&$1.10\pm0.03$&$0.88\pm0.02$&$0.36\pm0.02$\\
$N_{xillverCp}$($10^{-4}$)&$<0.54$&$12\pm3$&$12^{+4}_{-3}$&$9.7\pm0.6$&$4.8\pm0.4$&$2.0\pm0.2$\\
$C_{XRT}$&$1.064^{+0.008}_{-0.011}$&$1.013^{+0.015}_{-0.009}$&$1.082^{+0.014}_{-0.015}$&$1.052^{+0.025}_{-0.024}$&$1.042^{+0.024}_{-0.023}$&$0.87\pm0.04$\\
\hline
$L/L_{edd}$ (\%)&2.0&1.8&1.7&1.2&1.0&0.5\\
\hline
$C-stat$&\multicolumn{6}{c}{$11591$}\\
$\chi^2/d.o.f.$&\multicolumn{6}{c}{$12584/10731=1.173$}\\
\hline
\end{tabular}}
\end{center}
\raggedright{\textbf{Notes.} \\Luminosity calculated using unabsorbed flux
between $0.1-300$~keV, assuming a distance of 8 kpc and a black hole mass of 10~
$M_{\odot}$.}
\end{table*}

\bibliographystyle{apj}
\bibliography{draft-v14}
\end{document}